\documentclass[journal=jctc,manuscript=article]{achemso}

\usepackage[version=3]{mhchem} 
\usepackage{amsmath,accents}
\usepackage{amsthm}  
\usepackage{physics}
\usepackage{braket}
\usepackage{hyperref}
\usepackage{pdfpages}
\SectionNumbersOn
\usepackage[labelsep=period]{caption}

\hypersetup{
          colorlinks=true, linktocpage=true, pdfstartpage=3, pdfstartview=FitV,
          breaklinks=true, pdfpagemode=UseNone, pageanchor=true, pdfpagemode=UseOutlines,%
          plainpages=false, bookmarksnumbered, bookmarksopen=true, bookmarksopenlevel=1,%
          hypertexnames=true, pdfhighlight=/O,
          urlcolor=brown, linkcolor=blue, citecolor=green
            }

\hypersetup{
    colorlinks,
    linkcolor=blue,
    citecolor=blue
}

\author{Riccardo Alessandro}
\affiliation[Perugia]
{Dipartimento di Chimica, Biologia e Biotecnologie, Universit\`{a} degli Studi di Perugia, Via Elce di Sotto, 8,06123, Perugia, Italy}
\author{Matteo Castagnola}
\affiliation[Trondheim]
{Department of Chemistry, Norwegian University of Science and Technology, 7491 Trondheim, Norway}
\author{Henrik Koch}
\affiliation[Trondheim]
{Department of Chemistry, Norwegian University of Science and Technology, 7491 Trondheim, Norway}
\author{Enrico Ronca}
\affiliation[Perugia]
{Dipartimento di Chimica, Biologia e Biotecnologie, Universit\`{a} degli Studi di Perugia, Via
Elce di Sotto, 8,06123, Perugia, Italy}
\email{enrico.ronca@unipg.it}

\title[QED-CASSCF]
  {A Complete Active Space Self-Consistent Field approach for molecules in QED environments}
\begin{document}

%
%
%
%
%

\begin{abstract}
\noindent
Multireference systems are usually challenging to investigate using ab-initio methods as they require an accurate description of static electron correlation. 
The urgency of developing similar approaches is even more pressing when molecules strongly interact with light in quantum-electrodynamics (QED) environments.
In fact, in this context, multireference effects might be induced or reduced by the presence of the field.
In this work, we extend the Complete Active Space Self-Consistent Field (CASSCF) approach to polaritonic systems. The method is tested on benchmark multireference problems and applied to investigate field-induced effects on the electronic structure of well-known multiconfigurational processes.
Strengths and limitations of the method have been thoroughly analyzed.
\end{abstract}

\section{Introduction}
\begin{figure}[ht!]
    \includegraphics[width=1\linewidth]{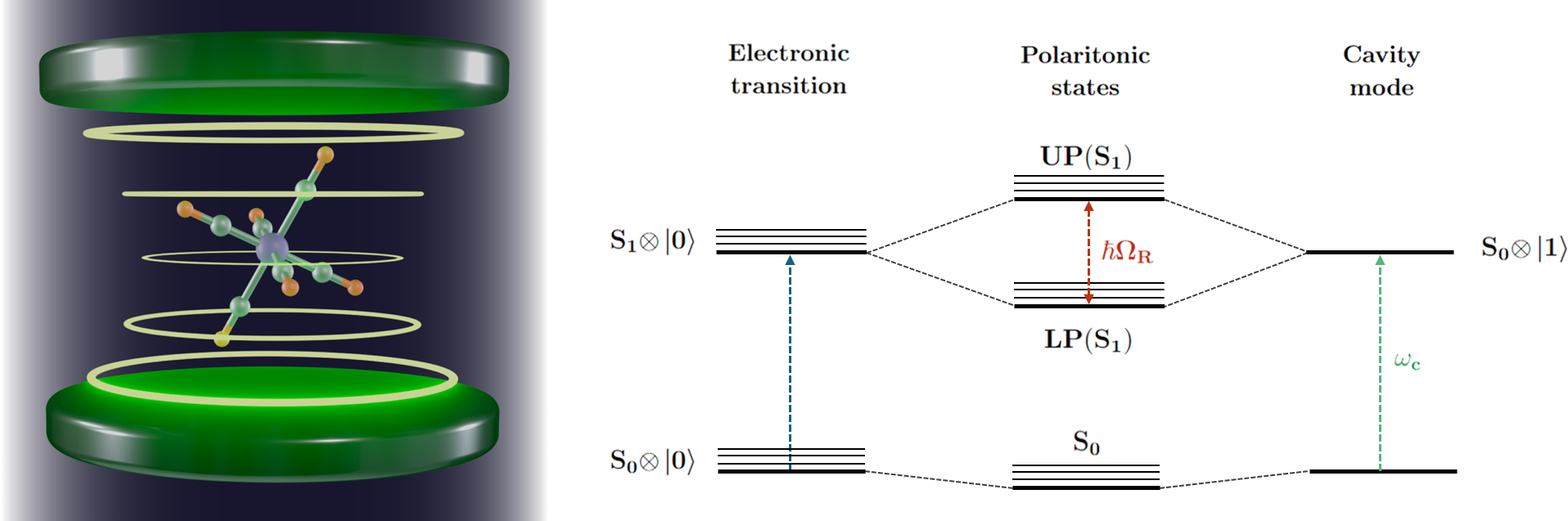}
    \caption{Illustration of a molecule in an optical cavity (left) and energy levels scheme of the polaritonic states obtained by coupling of an electronic transition of a molecule (left) and the cavity field (right). $\Omega_R$ indicates the Rabi splitting, while $\omega_c$ is the fundamental electromagnetic field frequency.}\label{fig:figure1}
\end{figure}

Manipulation of molecular properties by quantum fields is becoming nowadays a very promising non-intrusive alternative to chemical control~\cite{ebbesen2016hybrid, herrera2016cavity, galego2017many, martinez2018polariton, mandal2019investigating, wang2020coherent, delpo2020polariton, polak2020manipulating}. 
The observation of these effects was possible thanks to the significant technological advances made in the fabrication of efficient optical devices (optical~\cite{thompson1992observation, garcia2021manipulating}, plasmonic~\cite{baumberg2019extreme, baranov2020ultrastrong} and superconducting~\cite{teufel2011circuit, haroche2020cavity, wallraff2004strong} cavities, waveguides~\cite{sun2021polariton, downing2019topological, kondratyev2023probing}, etc.). 
The simplest of these devices is the Fabry–P\'{e}rot optical cavity (see Fig.~\ref{fig:figure1}), a device composed of two mirrors whose separation, L, is related to the fundamental wavelength of the field inside the cavity.
In these conditions, energy is coherently exchanged between the molecule and the cavity field, inducing a phenomenon called Rabi oscillations.
Due to the strong light-matter interaction, photons and molecular states lose their individuality and show mixed molecular and photonic character, generating hybrid states known as "polaritons"~\cite{miller2005trapped, flick2017atoms, feist2018polaritonic, hertzog2019strong}.
The formation of polaritons is associated with a splitting of the energy levels, known as \textit{Rabi splitting} ($\Omega_R$), that increases with the strength of the light-matter coupling ($\lambda$). 
It is noteworthy to point out that molecules can interact strongly with the field also in the absence of real photons, through a direct coupling between the molecular states and the vacuum fluctuations of the electromagnetic field.
In the past decade, a lot of research was devoted to exploring the potential of strong light-matter interaction on a plethora of physical or chemical phenomena. In particular, by coupling molecules or materials with different optical devices, it was possible to manipulate the absorption, emission, and scattering properties of molecules and materials~\cite{herrera2017absorption, baranov2020circular, shalabney2015enhanced, del2015signatures}, to induce Bose--Einstein condensation of polaritons~\cite{skolnick1998strong, kasprzak2006bose, byrnes2014exciton}, to increase conductivity in organic semiconductors~\cite{orgiu2015conductivity}, or to manipulate the spin properties of matter~\cite{bienfait2016controlling, bonizzoni2017coherent, bonizzoni2018coherent}.
One of the most debated but also interesting aspects was observed in chemistry, where field-induced effects on chemical reactivity were demonstrated to inhibit, catalyze, and even make some reactions selective toward specific products~\cite{ebbesen2016ground, ebbesen2019cavity, thomas2019tilting, sau2021modifying}.

Despite the large attention devoted to this field in the past decade, the physics behind many of these processes still remain unclear, and progress is still slow due to the many limitations in the validation of the experimental setup~\cite{ahn2023modification, patrahau2024direct, michon2024impact}. For these reasons, the development of effective theoretical approaches is crucial to overcome experimental limitations and to propose a consistent description of these phenomena~\cite{fregoni2022theoretical, foley2023ab}.
In recent years, a small number of research groups started developing some extensions to the most commonly used ab-initio methodologies, i.e., Density Functional Theory (DFT)~\cite{ruggenthaler2014quantum, ruggenthaler2018quantum, flick2018ab}, Hartree-Fock (HF), Full Configuration Interaction (FCI), Coupled Cluster (CC)~\cite{haugland2020coupled, mordovina2020polaritonic, pavosevic2021polaritonic, liebenthal2022equation, monzel2024diagrams}, and second-order M{\o}ller-Plesset perturbation theory (MP2)~\cite{bauer2023perturbation, moutaoukal2025strong}, to investigate coupled electron-photon systems. 
Recently, fully relativistic versions of QED-HF and QEDFT have also been proposed to investigate molecular systems containing heavy atoms in strong coupling conditions~\cite{thiam2024comprehensive,konecny2024relativistic}.
The quantum electrodynamics (QED) extensions to HF and DFT are quite efficient and can be easily applied to relatively large systems. However, they completely miss correlation or describe it in a very approximate way. Moreover, in the DFT case, an effective exchange-correlation functional for the electron-photon interaction is still far from being formulated~\cite{fregoni2022theoretical}. 

Coupled Cluster theory is known to be the most reliable method in quantum chemistry to simulate molecular systems and also for interacting electron-photon systems, its QED extension is at the moment the reference ab-initio methodology.
However, as for the standard electronic theory, the applicability of this method is limited to small- to medium-sized molecules, and its description becomes qualitatively poor when multiconfigurational effects come into play.
This problem is well-known and it prevents the investigation of a relatively large class of complexes, including several systems involved in chemically and biologically interesting processes~\cite{choudhury2024understanding}. 
For example, relevant molecular systems that can hardly be described with CC methodologies include open-shell transition metal complexes, (poly)radicalic systems, excited states of organic molecules, and torsion or breakings of conjugated bonds~\cite{lischka2018multireference, szalay2012multiconfiguration}.

The state-of-the-art reference approaches usually used to investigate these systems are the Multiconfigurational Self-Consistent Field (MCSCF) methods, which are based on a wavefunction defined as a linear combination of a reduced number of determinants or Configuration State Functions (CSFs). 
A wide variety of MCSCF methods have been developed. The main difference between them lies in the protocol employed to select the determinants to include in the expansion~\cite{helgaker2013molecular}. 
The most popular MCSCF method is the Complete Active Space Self-Consistent Field (CASSCF), which is based on a partitioning of the orbital space into three classes: \textit{inactive}, \textit{virtual}, and \textit{active} orbitals.
The multideterminantal expansion in the case of CASSCF theory is obtained by choosing a certain number of so-called \textit{active electrons} and computing all the possible excitations of these electrons within the active space. If the active space is chosen properly (i.e., the most correlated orbitals and electrons are included), the CASSCF wavefunction is able to provide a qualitatively-correct description of the multireference molecular system. 
Despite this method being equivalent to a FCI expansion within the active space, with the current computational facilities, it can be applied, in a relatively efficient way, to active spaces with up to 20 electrons in 20 orbitals~\cite{vogiatzis2017pushing}.

In the latest years, some of the most common ab-initio multireference methods have been extended to coupled light-matter systems~\cite{qed_casci, qed_dmrg, qed_rmd2}. In particular, the implementation of QED-CASCI and QED-DMRG provided the first steps towards the inclusion of static correlation effects in optical cavities. 
However, the CASCI procedure does not optimize the orbitals, leading to variationally higher energies with respect to its SCF counterpart~\cite{levine2021cas}, and although the DMRG method has a polynomial scaling, its cost is higher than CASSCF for small active spaces. 
The comparatively higher computational cost of DMRG for small active spaces provides a rationale for extending and applying the CASSCF method to the QED case.

In this work, we present an extension of the State-Specific (SS-)CASSCF theory to coupled light-matter systems (QED-CASSCF). 
The implementation is performed following a restricted-step second-order optimization scheme, according to the implementations in Refs.~\citenum{jensen1984direct, hoyvik2012trust, folkestad2022implementation, helmich2022trust}. 
Recently, a QED version of the State-Averaged (SA) variant of CASSCF was proposed by Vu et al.~\cite{vu2025cavity} in parallel to this work. It was applied to a small test case with an active space close to the full orbital space, with results very close to the FCI limit.
In this work, we perform simulations of larger systems where the CASSCF method does not approach the FCI limit.
In this situation, a deep analysis of the method's performances to describe field-induced effects on multireference systems could be performed.

This paper is organized as follows: in Section~\ref{section:2}, a brief summary of the CASSCF method and its second-order implementation is presented. 
In Section~\ref{section:3}, the fundamental details of QED-CASSCF are described. 
In Section~\ref{section:4}, QED-CASSCF is tested on small molecules against other ab-initio methodologies. 
In this section, QED-CASSCF is also used to investigate the field effects on well-known multireference chemical processes. 
We conclude the work with future perspectives presented in Section~\ref{section:5}.
\section{CASSCF Theory \label{section:2}}
\begin{figure}[H]
    \centering
    \includegraphics[width=0.5
    \linewidth]{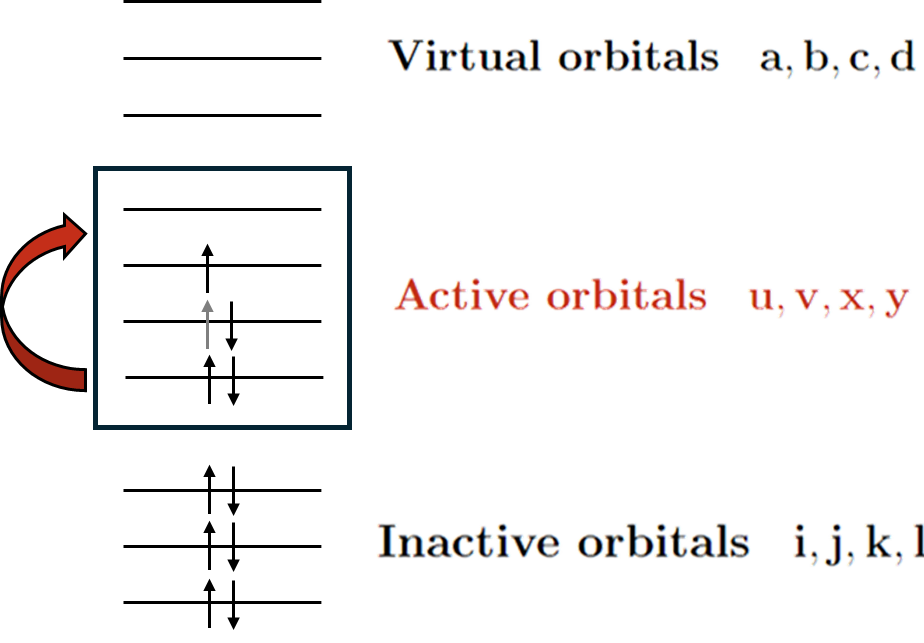}
    \caption{CASSCF orbital partition scheme.}\label{fig:figure2}
\end{figure}

In MCSCF theory the wavefunction is written as the linear combination of a certain number of Slater determinants or Configuration State Functions (CSFs)
\begin{equation}
    \ket{\Psi^{\text{MCSCF}}} = \sum_I C_I \ket{I}
\end{equation}
and it is optimized by variationally minimizing the energy with respect to the molecular orbital-rotation parameters and configuration interaction (CI) coefficients ($\boldsymbol{\kappa}$ and $\boldsymbol{c}$, respectively)
\begin{equation}
    E = \min_{\boldsymbol{\kappa}, \mathbf{c}} \frac{\matrixel{\Psi}{\hat{H}}{\Psi}}{\braket{\Psi | \Psi}} .
\end{equation}
Among the MCSCF methods~\cite{olsen1988determinant, malmqvist1990restricted, fleig2001generalized, schollwock2005density, mazziotti2006quantum}, the Complete Active Space Self-Consistent Field (CASSCF) approach is still the state-of-the-art procedure for molecular systems dominated by static correlation. The orbital space  of a molecule in the CASSCF approximation is partitioned into three subspaces, as depicted in Fig.~\ref{fig:figure2}: inactive orbitals are always fully occupied, active orbitals have non-integer occupations, and virtual orbitals are always empty. 
In this work the different orbitals are addressed according to the following indices choice: $i,j,k,l$ for inactive orbitals, $u,v,x,y$ for active orbitals, $a,b,c,d$ for virtual orbitals, and $p,q,r,s$ for generic orbitals.
The multideterminantal expansion for the wavefunction is obtained by considering all the determinants computed by distributing $N$ active electrons in $M$ active orbitals, namely CAS(N,M).

The first implementations of CASSCF relied on first-order solvers, and among these, the super-CI approach is the most widely applied~\cite{roos1980sci1, roos1980sci2}. 
Although recent first-order implementations proved to be very efficient~\cite{superCI_PT}, during the optimization of the CASSCF wavefunction, convergence issues may arise. 
\color{black}
This led to the development of robust second-order quadratically-convergent solvers, which exhibit better convergence properties in a reduced number of iterations at the price of a higher cost for each step~\cite{jensen1984direct, jensen1986direct, jensen1987efficient, werner1980quadratically, werner1985second, sun2017general, ma2017second, kreplin2019second, kreplin2020mcscf, nottoli2021second, helmich2022trust}. 
CASSCF implementations based on second-order algorithms are nowadays relatively well-established in the quantum chemistry community and are available in the majority of the most used software packages~\cite{aidas2014d, aquilante2020modern, matthews2020coupled, sun2020recent, neese2012orca}. 
This approach is particularly convenient for the application to simulate small molecular systems, and calculations on larger systems are nowadays efficiently performed with the aid of parallelization, AO-based implementations, or by exploiting the Cholesky decomposition or the resolution of identity (RI) approximation of the two-electron integrals~\cite{nottoli2021second, helmich2022trust}.

In MCSCF theory, the first step towards the implementation of the method is an appropriate choice of the parametrization for the wavefunction. 
The orbital part is parametrized in terms of an exponential transformation with the anti-Hermitian operator $\hat{\kappa}$ defined as:
\begin{equation}
    \hat{\kappa} = \sum_{p>q} k_{pq} \hat{E}_{pq}^{-} = 
                   \sum_{p>q} k_{pq} (\hat{E}_{pq} - \hat{E}_{qp}),
\end{equation}
where $\hat{E}_{pq} = \sum_{\sigma} \hat{a}^{\dag}_{p\sigma} \hat{a}_{q\sigma}$ is the spin traced singlet excitation operator with $\sigma \in \{\alpha, \beta\}$. 
While no alternatives to the parametrization of the orbital space are available, the CI space is liable to multiple choices depending on the CASSCF variant that has to be implemented. 
In the case of State-Specific (SS) CASSCF method, a linear parametrization for the CI part is commonly adopted and the wavefunction can be written as~\cite{jensen1984direct}:
\begin{equation}
    \ket{\Psi} = e^{-\hat{\kappa}}\frac{\ket{0} + \hat{P}\ket{\mathbf{c}}}{\sqrt{1 + \matrixel{\mathbf{c}}{\hat{P}}{\mathbf{c}}}}
\end{equation} 
where $\ket{0}$ is indicated as the \textit{current expansion point} (CEP) and constitutes the current approximation to the wavefunction. $\hat{P} = 1 - \ket{0} \! \bra{0}$, instead, is a projector operator that removes the components parallel to the CEP from the correction vector, $\ket{\mathbf{c}}$, which contains the variational parameters, $c_I$, with respect to the CI part:
\begin{equation}
    \ket{\mathbf{c}} = \sum_I c_I \ket{I} .
\end{equation}\color{black}
The choice of the exponential parametrization allows to consider only the non-redundant rotations between the orbital spaces (i.e., orbital rotations that contribute to the energy: inactive-active, inactive-virtual, and active-virtual)~\cite{helgaker2013molecular}. 
The linear parametrization of the CI space contains only one redundant parameter, that is, when the correction vector $\ket{\mathbf{c}}$ has components parallel to the reference state $\ket{0}$. However, this is not an issue, as this redundancy can be easily tracked and is projected out by the operator $\hat{P}$. Therefore, this redundancy does not interfere with the optimization procedure.

In the following of this section, we will quickly recap the implementation details of the second-order SS-CASSCF algorithm that will be applied, in the next section, for the extension to strongly coupled electron-photon systems.

\subsection{Second-Order CASSCF}
The implementation of a second-order algorithm is based on the definition of a quadratic model to locally expand the energy up to the second order in $\boldsymbol{\delta}$:
\begin{equation}\label{quadratic}
    E \approx Q(\boldsymbol{\delta}) = E_0 + \mathbf{g}^T \boldsymbol{\delta} + \frac{1}{2} \boldsymbol{\delta}^{{T}} \mathbf{G} \boldsymbol{\delta}
\end{equation}
where $\boldsymbol{\delta}$ indicates a small variation of the wavefunction parameters ($\boldsymbol{\kappa}$, $\boldsymbol{c}$). 
In Eq.~\ref{quadratic}, $E_0$ is the CASSCF energy and $\mathbf{g}$ is the gradient vector containing the first derivatives of the energy with respect to the configuration (c) and orbital (o) coefficients, all calculated at the current expansion point:
\begin{equation}\label{gradient}
    \mathbf{g} = 
    \begin{pmatrix}
        \mathbf{g}^c \\ \\
        \mathbf{g}^o 
    \end{pmatrix}
    =
    \begin{pmatrix}
        \dfrac{\partial E}{\partial \mathbf{c}} \\ \\
        \dfrac{\partial E}{\partial \boldsymbol{\kappa}} .
    \end{pmatrix}
\end{equation}
Finally, $\mathbf{G}$ is the Hessian matrix containing the pure configuration (cc), orbital (oo), and the mixed configuration-orbital (co) and orbital-configuration (oc) parts:
\begin{equation}\label{hessian}
        \mathbf{G} =
        \begin{pmatrix}
            \mathbf{G}^{cc}  & \mathbf{G}^{co} \\ \\
            \mathbf{G}^{oc}  & \mathbf{G}^{oo}
        \end{pmatrix}
        =
        \begin{pmatrix}
            \dfrac{\partial ^ 2 E}{\partial \mathbf{c} \partial \mathbf{c}}         & 
            \dfrac{\partial ^ 2 E}{\partial \mathbf{c} \partial \boldsymbol{\kappa}} \\ \\
            \dfrac{\partial ^ 2 E}{\partial \boldsymbol{\kappa} \partial \mathbf{c}} & 
            \dfrac{\partial ^ 2 E}{\partial \boldsymbol{\kappa} \partial \boldsymbol{\kappa}}
        \end{pmatrix} . 
\end{equation}
The Hessian matrix elements are calculated, again, at the current expansion point.

In principle, a second order algorithm requires the explicit computation of the electronic Hessian. However, this procedure can be performed efficiently in a direct fashion. This approach passes through the calculation of the so-called \textit{sigma vectors} defined as:
\begin{equation}\label{sigma_vectors}
    \boldsymbol{\sigma} = \mathbf{G} \mathbf{b} = 
    \begin{pmatrix}
        \mathbf{G}^{cc}  & \mathbf{G}^{co} \\ \\
        \mathbf{G}^{oc}  & \mathbf{G}^{oo}
    \end{pmatrix}
    \begin{pmatrix}
        \mathbf{b}^c \\ \\
        \mathbf{b}^o
    \end{pmatrix}
    =
    \begin{pmatrix}
        \mathbf{G}^{cc} \; \mathbf{b}^c + \mathbf{G}^{co} \; \mathbf{b}^o \\ \\
        \mathbf{G}^{oc} \; \mathbf{b}^c + \mathbf{G}^{oo} \; \mathbf{b}^o
    \end{pmatrix}.
\end{equation}
Here, $\mathbf{b}$ are the so-called trial vectors used in the Davidson scheme~\cite{davidson197514} which contain the wavefunction parameters with respect to the orbital and configuration parts
\begin{equation}
    b^o _{pq} = \kappa_{pq} , \qquad\qquad b^c _I = v_I .
\end{equation}
This approach allows for large scale operations without explicitly calculating and storing the blocks of $\mathbf{G}$ in memory~\cite{roos1972new}.
Moreover, sigma vectors can be conveniently computed applying minor modifications to intermediate quantities already calculated for the gradients.

By minimizing the quadratic function defined in Eq.~\ref{quadratic} the equation for the Newton step is obtained as:
\begin{equation}\label{newton_step}
     \mathbf{G} \boldsymbol{\delta} = - \mathbf{g} .
\end{equation}
This procedure is not optimal since it requires the Hessian to be positive-definite. This condition is not always guaranteed, especially at the beginning of the optimization procedure. Even when it is respected, the computed step may not point in the direction of the minimum, leading to a poor convergence~\cite{fletcher2000practical}.
This problem can be largely reduced by adopting a restricted step procedure~\cite{jensen1984direct, simons1983walking}. According to this scheme, the choice of the step is restricted within a well-defined \textit{trust region} with radius, $h$:
\begin{equation}
    \boldsymbol{\delta}^T \boldsymbol{\delta} \leq h
\end{equation}
where we assume the potential energy surface to be correctly approximated by the quadratic model. This ensures the Hessian to be positive-definite. 
This procedure was first described by Levenberg and Marquardt~\cite{levenberg1944method, marquardt1963algorithm} and is based on the solution of the level-shifted Newton equation:
\begin{equation}\label{LM_eq}
    (\mathbf{G} - \mu \mathbf{I}) \boldsymbol{\delta} = - \mathbf{g}
\end{equation}
where the level-shifting parameter $\mu$ corresponds to the lowest eigenvalue of the \textit{augmented Hessian}:
\begin{equation}\label{augmented_hessian}
    \mathbf{L}(\alpha) =
    \begin{pmatrix}
        0 & \alpha\mathbf{g}^{T} \\
        \alpha\mathbf{g} & \mathbf{G}
    \end{pmatrix}
    .
\end{equation}
The relation between the eigenvalues $\epsilon$ of $\mathbf{G}$ and $\mu$ of $\mathbf{L}(\alpha)$ \color{black} is given by the Hylleraas-Undheim-MacDonald theorem~\cite{hylleraas1930numerische, macdonald1933successive}:
\begin{equation}
    \mu_1 \leq \epsilon_1 \leq \mu_2 \leq \dots \leq \mu_n \leq \epsilon_n \leq \mu_{n+1} .
\end{equation}
Therefore, the choice of the lowest eigenvalue of the augmented Hessian as a level-shifting parameter guarantees $(\mathbf{G} - \mu \mathbf{I})$ to be positive-definite.
This approach requires, in principle, to solve the linear system in Eq.~\ref{LM_eq} at every iteration. 
However, the inversion of $(\mathbf{G} - \mu \mathbf{I})$ could be computationally inefficient.
An alternative way can be obtained by diagonalizing the augmented Hessian in Eq.~\ref{augmented_hessian}.
This can be done efficiently in an iterative fashion by implementing the sigma vectors in Eq.~\ref{sigma_vectors}. Once the lowest eigenvalue and eigenvector are obtained, the step for the new iteration can be computed as:
\begin{equation}\label{newstep}
    \boldsymbol{\delta} = \frac{1}{\alpha} \mathbf{x}(\alpha)
\end{equation}
where $\mathbf{x}(\alpha)$ is the eigenvector of the Hessian and $\alpha$ is a scaling parameter that forces the step to lie withing the trust region.
The trust radius is changed adaptively during the procedure following the strategy described by Fletcher in Ref.~\citenum{fletcher2000practical}. As long as the step is kept within the trust region, the Newton's step in ~\ref{newton_step} is recovered, and convergence to the closest minimum is guaranteed.

The presented trust-region optimization method can be summarized as follows:
\begin{enumerate}
    \item Compute the gradient in Eq.~\ref{gradient}. If its norm is lower than a certain threshold, convergence is reached;
    \item Check the step and adjust the trust radius according to the procedure described by Fletcher in Ref.~\citenum{fletcher2000practical};
    \item Iteratively diagonalize the augmented Hessian in Eq.~\ref{augmented_hessian};
    \item Update the wavefunction parameters (Eq.~\ref{newstep}) and return to step 1.
\end{enumerate}

Details on this procedure are described in Refs.~\citenum{jensen1984direct, jensen1987efficient, fletcher2000practical, helgaker2013molecular}. The optimized algorithm as implemented in the e$^{\mathcal T}$ suite of programs~\cite{eT}, is described in Ref.\citenum{folkestad2022implementation}.
\section{QED-CASSCF \label{section:3}}
The description of molecular systems in optical cavities requires to account explicitly for the quantum character of the electromagnetic field. In our work, we will use the well-established non-relativistic single-mode Pauli-Fierz Hamiltonian in length gauge and dipole approximation as a starting point for the development of the QED-CASSCF approach~\cite{haugland2020coupled}:
\begin{equation}\label{pf_original}
    \hat{H}_{\text{PF}} =  \hat{H}_e +
    \omega \hat{b}^{\dag}\hat{b} + 
    \frac{1}{2}(\boldsymbol{\lambda} \cdot (\boldsymbol{\hat{d}} - \braket{\boldsymbol{\hat{d}}} ))^2 - 
    \sqrt{\frac{\omega}{2}}(\boldsymbol{\lambda} \cdot (\boldsymbol{\hat{d}} - \braket{\boldsymbol{\hat{d}}} ))(\hat{b}^{\dag} + \hat{b}) .
\end{equation}
Here, $\boldsymbol{\hat{d}}$ is the dipole moment operator, $\braket{\boldsymbol{\hat{d}}}$ is its expectation value, and $\boldsymbol{\lambda} = \sqrt{\frac{4\pi}{V}} \boldsymbol{e} $ is the light-matter coupling parameter that depends on the quantization volume $V$ and the polarization vector $\boldsymbol{e}$. 
For simplicity, a single cavity mode is included in our treatment, but extension to more modes can be applied in a trivial manner. 
To ensure origin invariance, the Hamiltonian in Eq.~\ref{pf_original} has already been expressed in the coherent-state basis~\cite{haugland2020coupled}. Following the notation used in Ref.~\citenum{lexander2024analytical}, we can conveniently set $\hat{d} = \boldsymbol{\hat{d}} \cdot \boldsymbol{\lambda}$ and rewrite the Pauli-Fierz Hamiltonian as
\begin{equation}
    \begin{split}
        \hat{H}_{\text{PF}} &= \sum_{pq} h_{pq} \hat{E}_{pq} + 
        \frac{1}{2} \sum_{pqrs} g_{pqrs} \hat{e}_{pqrs} + 
        \omega \hat{b}^{\dag}\hat{b} \\
        &+ \sqrt{\frac{\omega}{2}} \sum_{pq}d_{pq}\hat{E}_{pq} (\hat{b}^{\dag} + \hat{b}) - 
        \sqrt{\frac{\omega}{2}} \braket{\hat{d}} \! (\hat{b}^{\dag} + \hat{b}) + \hat{h}_{nuc} .
    \end{split}
\end{equation}
In this form the dipole-self energy term is included in the one- and two-electron integrals
\begin{equation}
    \begin{split}
        & h_{pq} = h^e _{pq} + \frac{1}{2} q_{pq} - d_{pq} \! \braket{\hat{d}} + \frac{\delta_{pq}}{2N_e} \! \braket{\hat{d}} ^2 \\
        & g_{pqrs} = g^e _{pqrs} + d_{pq}d_{rs}
    \end{split}
\end{equation}
with $d_{pq}$ and $q_{pq}$ being the electric dipole and quadrupole integrals, respectively, and $N_e$ is the number of electrons in the molecule.

For a general QED-MCSCF theory, the ground-state \color{black} wavefunction can be expressed as~\cite{qed_casci}: 
\begin{equation}
    \ket{\Psi_0} = \sum_I \sum_m C_{I,m}^{(0)} \ket{I} \otimes \ket{m}
\end{equation}
where $\ket{m}$ is the $m^{th}$ photonic state of the cavity mode. In this case the CI coefficients refer to the non-interacting electron-photon states $\Phi_{I,m}$ defined as
\begin{equation}
    \Phi_{I,m} = \ket{I} \otimes \ket{m} .
\end{equation}

For the sake of readability of the following section, we conveniently separate the Pauli-Fierz Hamiltonian into three terms:
\begin{equation}
       \hat{H}_{\text{PF}} = \hat{H}_e + \hat{H}_{ph} + \hat{H}_{int}
\end{equation}
where we need to remember that $H_e$ contains the one- and two-electron integrals modified for the dipole self-energy term.
From now on, we will refer to the Pauli-Fierz Hamiltonian as ${H}$ unless explicitly stated.

The extension of CASSCF to molecular polaritons and its second-order implementation requires the computation of modified energy contributions, gradients, and sigma vectors. The expression for the QED-CASSCF energy can be written as:
\begin{equation}
    E^{\text{QED-CASSCF}} = \matrixel{\Psi_0}{\hat{H}}{\Psi_0} = E_{el} + E_{ph} + E_{int} + E_{nuc}
\end{equation}
where
\begin{equation}\label{electronic_energy}
    E_{el} = \sum_{uv} F^I_{uv} \gamma_{uv} + 
        \frac{1}{2} \sum_{uxyv} g_{uvxy} \Gamma_{uvxy} + 
        \sum_i \left( h_{ii} + F^I _{ii} \right) .
\end{equation}
\begin{equation}
    \gamma_{uv} = \matrixel{\Psi_0}{\hat{E}_{uv}}{\Psi_0}, \quad\quad 
    \Gamma_{uvxy} = \matrixel{\Psi_0}{\hat{e}_{uvxy}}{\Psi_0}
\end{equation}
Are the one- and two-body reduced density matrices, respectively, where
\begin{equation}
    \hat{e}_{uvxy} = \hat{E}_{uv}\hat{E}_{xy} - \delta_{xv} \hat{E}_{uy}   
\end{equation}
is the two-electron excitation operator running on the active indices~\cite{helgaker2013molecular}.
In Eq.~\ref{electronic_energy}
\begin{equation}
    F_{pq}^I = h_{pq} + \sum_i (2g_{pqii} - g_{piiq})
\end{equation}
are the elements of the inactive Fock matrix~\cite{jensen1986direct, helgaker2013molecular}, and 
\begin{equation}
    E_i = \sum_i \left( h_{ii} + F^I _{ii} \right) .
\end{equation}
is denoted as inactive energy~\cite{nottoli2021second}.
The purely photonic contribution to the energy is given by:
\begin{equation}
    E_{ph} = \omega \sum_{I,m} |C^{(0)}_{I,m}|^2 m
\end{equation}
where $m$ is the number of photons in the $m^{th}$ photonic state of the cavity mode.
Finally, the contribution by the bilinear coupling is written as:
\begin{equation}
    E_{int} = \sqrt{\frac{\omega}{2}} \Bigg\{ \Big(
    \sum_i 2d_{ii} - \braket{\hat{d}} \Big) \sum_{I,m} \left( 
    \sqrt{m + 1} C^{(0)}_{I,m+1} + \sqrt{m} C^{(0)}_{I,m-1} 
    \right) C^{(0)}_{I,m}
    + \sum_{uv} d_{uv} \Tilde{\gamma}_{uv} \Bigg\}
\end{equation}
where the modified one-body density matrix is defined as:
\begin{equation}
     \Tilde{\gamma}_{uv} = \sum_{I,J} \sum_m \left(
     \sqrt{m+1} C^{(0)}_{I,m+1} \matrixel{I}{\hat{E}_{uv}}{J} C^{(0)}_{J,m} + 
     \sqrt{m} C^{(0)}_{I,m-1} \matrixel{I}{\hat{E}_{uv}}{J} C^{(0)}_{J,m} 
     \right) .
\end{equation}
The general expression for the CI gradient is given by:
\begin{equation}
    g_{I,m} ^c = 2\matrixel{\Phi_{I,m}}{\hat{P}\hat{H}}{\Psi_0} = 2\matrixel{\Phi_{I,m}}{\hat{H}}{\Psi_0} - 2C_{I,m} ^{(0)}E^{(0)}
\end{equation}
where $E^{(0)}$ is the QED-CASSCF energy at the current expansion point.
Similarly to the energy, also the CI gradient can be split up into different terms:
\begin{equation}
        g_{I,m} ^c = 2\left( g^{el}_{I,m} + g^{ph}_{I,m} + g^{int}_{I,m} - C_{I,m}^{(0)}E^{(0)} \right)
\end{equation}
where
\begin{equation}
    \begin{split} 
        g^{el}_{I,m} &= \matrixel{\Phi_{I,m}}{\hat{H}^{el}}{\Psi_0} \\
        &= E_i C^{(0)}_{I,m} + \sum_{J}  
        \matrixel{I}{\sum_{uv} F^I _{uv} \hat{E}_{uv} +
        \frac{1}{2} \sum_{uvxy} g_{uvxy} \hat{e}_{uvxy}}{J} 
        C_{J,m} ^{(0)} 
    \end{split}
\end{equation}
and
\begin{equation}
    g^{ph}_{I,m} = \matrixel{\Phi_{I,m}}{\hat{H}_{ph}}{\Psi_0} = m \omega C_{I,m}^{(0)} .
\end{equation}
The bilinear contribution to the CI gradient:
\begin{equation}
    \begin{split}
        g^{int}_{I,m} &= \matrixel{\Phi_{I,m}}{\hat{H}_{int}}{\Psi_0} \\
        &= \sqrt{\frac{\omega}{2}} \sum_n \left( 
        \sqrt{n}\delta_{m,n-1} + \sqrt{n+1}\delta_{m,n+1} 
        \right) \Bigg\{ 
        \Big( \sum_i 2d_{ii} - \braket{\hat{d}} \Big) C_{I,n}^{(0)} +
        \sum_J \sum_{uv} d_{uv} \matrixel{I}{\hat{E}_{uv}}{J}
        C_{J,n} ^{(0)} \Bigg\}
    \end{split}
\end{equation}
The general expression for the orbital part of the gradient from Eq.~\ref{gradient} is given by:
\begin{equation}
    g_{mn} ^o = \matrixel{\Psi_0}{[\hat{E}_{mn}^- , \hat{H}]}{\Psi_0} = 2 \matrixel{\Psi_0}{[\hat{E}_{mn} , \hat{H}]}{\Psi_0}
\end{equation}
where the purely photonic contribution to the orbital gradient is zero since $\hat{E}_{mn}$ and $\hat{H}_{ph}$ commute.
The purely electronic contribution is known~\cite{helgaker2013molecular}:
\begin{equation}
    g_{mn} ^{el} = 2(F_{mn} - F_{nm})
\end{equation}
where 
\begin{equation}
    F_{mn} = \sum_q \gamma_{mq} h_{nq} + \sum_{qrs} \Gamma_{mqrs} g_{nqrs} .
\end{equation}
is the generalized Fock matrix.
Finally, the interaction term to the gradient reads as:
\begin{equation}\label{bilinear_orbital_gradient}
        g_{pq}^{int} = 2\matrixel{\Psi_0}{[\hat{E} _{pq}, \hat{H}^{int}]}{\Psi_0}
%
%
%
        = 2 \sqrt{\frac{\omega}{2}} \sum_r \left( d_{qr} \Tilde{\gamma}_{pr} - 
        d_{rp} \Tilde{\gamma}_{rq} \right) .
\end{equation}
The sigma vectors defined in Eq.~\ref{sigma_vectors} can be calculated as: 
\begin{subequations}
    \begin{align}
            \sigma^{cc}_{I,m} &= \sum_{J,n} G^{cc}_{I,m;J,n} \; b^c_{J,n} = 
            2 \matrixel{\Phi_{I,m}}{(\hat{H} - E_0)}{\mathbf{v}} 
            - C_{I,m}^{(0)} \sum_{J,n} g_{J,n}^c v_{J,n} 
            - g_{I,m}^c \sum_{J,n} C_{J,n}^{(0)} v_{J,n} \\
            \sigma^{oc}_{pq} &= \sum_{I,m} G^{oc}_{pq;I,m} \; b^c_{I,m} = 
            \matrixel{\mathbf{v}}{[\hat{E}_{pq}^- , \hat{H}]}{\Psi_0} +
            \matrixel{\Psi_0}{[\hat{E}_{pq}^- , \hat{H}]}{\mathbf{v}} - 
            2 g^o_{pq} \sum_{I,m} C_{I,m} ^{(0)} v_{I,m} \\
            \sigma^{co}_{I,m} &= \sum_{pq} G^{co}_{I,m;pq} \; b^o_{pq} = 
            2\matrixel{\Phi_{I,m}}{\hat{H}^k}{\Psi_0} - 2 C_{I,m}^{(0)} \sum_{pq} g^o_{pq}\kappa_{rs} \\
            \sigma^{oo}_{pq} &= \sum_{rs} G_{pq,rs} \; b^o_{rs} = \matrixel{\Psi_0}{[\hat{E}_{pq} ^- , \hat{H}^k]}{\Psi_0} + \frac{1}{2}
            \sum_s (g^o_{sp} \kappa_{qs} - g^o_{sq} \kappa_{ps})
    \end{align} 
\end{subequations}
where we have introduced the modified state
\begin{equation}
    \ket{\mathbf{v}} = \sum_{I}\sum_{m} v_{I,m} \ket{\Phi_{I,m}} .
\end{equation}
$\hat{H}^k$ is the one-index transformed Hamiltonian~\cite{helgaker2013molecular}, where only the terms involving molecular integrals are one-index transformed:
\begin{equation}
    \hat{H}^k_{el} = \sum_{pq} h^k_{pq} \hat{E}_{pq} + \frac{1}{2} \sum_{pqrs} g^k_{pqrs} \hat{e}_{pqrs}
\end{equation}
\begin{equation}
    \hat{H}^k_{int} = \sqrt{\frac{\omega}{2}} \sum_{pq} d^k_{pq} \hat{E}_{pq} (\hat{b}^{\dag} + \hat{b})
\end{equation}
where 
\begin{equation}
    h^k_{pq} = \sum_m (k_{pm}h_{mq} + k_{qm}h_{pm})
\end{equation}
\begin{equation}
    g^k_{pqrs} = \sum_m (k_{pm}g_{mqrs} + 
                          k_{qm}g_{pmrs} +
                          k_{rm}g_{pqms} +
                          k_{sm}g_{pqrm})
\end{equation}
\begin{equation}
    d^k_{pq} = \sum_m (k_{pm}d_{mq} + k_{qm}d_{pm})
\end{equation}
Substituting these quantities in Eqs.~\ref{gradient}-\ref{sigma_vectors} an efficient second-order algorithm for QED-CASSCF can be naturally obtained.
\section{Results \label{section:4}} 
In this section, QED-CASSCF will be tested on small multireference systems and used to investigate field-induced multiconfigurational effects. All the calculations are performed with a development version of the $e^\mathcal{T}$ software package~\cite{eT}.
\subsection{Dissociation of the N$_2$ ground state}
\begin{figure}[H]
    \centering
    \includegraphics[width=\linewidth]{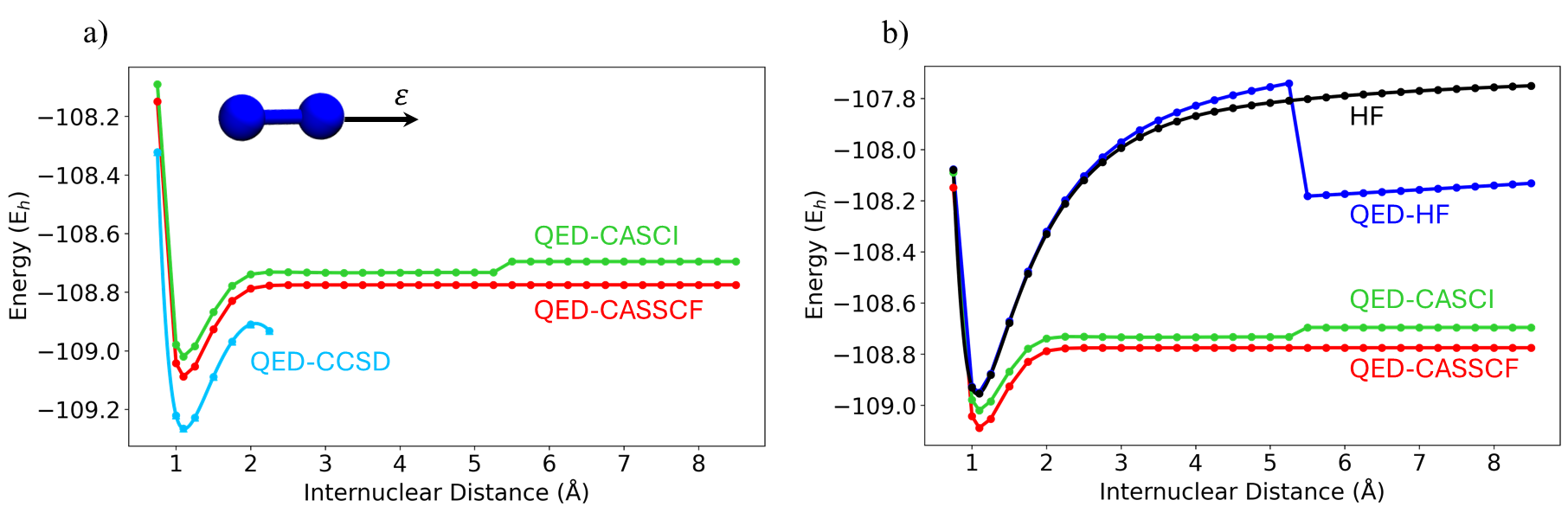}
    \caption{Potential energy curves of the nitrogen molecule along the internuclear distance computed at the \textbf{a.} QED-CASSCF, QED-CASCI, and QED-CCSD level \textbf{b.} QED-CASSCF, QED-CASCI, QED-HF, and purely electronic HF.}  
    \label{fig:figure3}
\end{figure}

The homolytic dissociation curve of N$_2$ is a typical test case for investigating the behavior of multireference methods. 
For this system, the active space contains the six valence electrons inside the six p orbitals, i.e., CAS(6,6).
The calculations were performed using a Dunning's cc-pVDZ basis~\cite{dunning1989gaussian}, light-matter coupling $\lambda= 0.03$ a.u., cavity frequency $\omega=0.5$ eV, and the field polarization along the $z$ axis, coincident with the main molecular axis.
All the QED calculations are performed including a single photonic state. To check the potential effect of a higher number of photons in the photonic Fock space, the calculations were repeated including up to three photonic states, but no relevant difference was observed (see Fig.~S1 in the Supporting Information).
The potential energy curve has been computed over 32 steps from $0.75$ to $8.50${\AA}.
In Fig.~\ref{fig:figure3}a, the potential energy curve computed at the QED-CASSCF level is compared against QED-CASCI and QED-CCSD.
For distances up to 5{\AA}, the energy profiles are in qualitative agreement with the purely electronic potential energy curves (PECs), with coupled cluster exhibiting lower energies with respect to CAS-type energies. Notice, however, that, at $2.25${\AA}, CC, after showing an unphysical maximum around $2.0${\AA}, stops converging due to the emergence of multireference effects associated with the homolytic dissociation of the system. 
Despite showing a higher energy for the dissociation plateau compared to both QED-CC and QED-CASSCF, QED-CASCI describes the right energy profile in a wide range of energies, as expected already for the bare electronic system. The higher energy of the dissociated system can be clearly explained by the lack of dynamical correlation (included in QED-CC) and by the use of unoptimized molecular orbitals (present in QED-CASSCF). 
However, it is important to highlight the presence of an unphysical jump in energy between $5.25$ and $5.50${\AA} which is not registered for the purely electronic profile (see Fig.~S2 in Supporting Information).
This behavior is indeed given by the guess used for the calculation. In fact, while the purely electronic Hartree-Fock potential energy curve is continuous in the whole energy range, the QED-HF curve shows a drastic energy change at the same internuclear distances (see Fig.~\ref{fig:figure3}b). 

\begin{figure}[H]
    \centering
    \includegraphics[width=\linewidth]{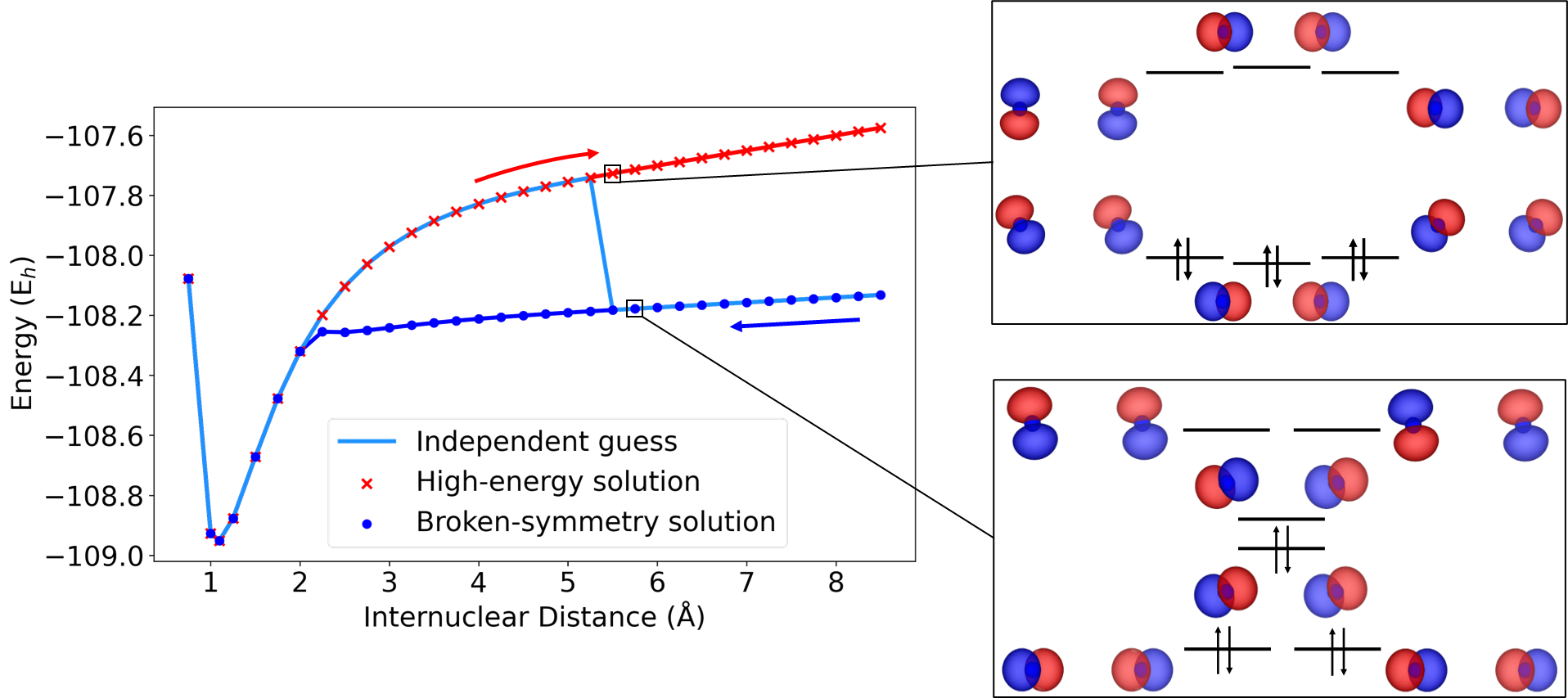}
    \caption{QED-HF potential energy curves for the nitrogen molecule obtained by restarting the calculations using the orbitals from lower to higher distances (in red) and from higher to lower distances (in blue), compared with the curve obtained by generating an independent guess for each point (in light blue).}  
    \label{fig:figure4}
\end{figure}

The QED-HF orbitals used in Fig.~\ref{fig:figure3} are generated independently at every point. To thoroughly understand the unphysical behavior at high distances, we performed the calculations at the QED-HF level to obtain the two full PECs by using different orbitals.
A higher-energy curve was obtained by restarting the calculations using the orbitals from smaller internuclear distances while a lower-energy curve was obtained by using the orbitals from larger distances (see Fig.~\ref{fig:figure4}).
The higher energy curve reproduces both the profile and the orbital energy ordering of the purely electronic HF dissociation curve while the lower energy curve corresponds to a symmetry broken solution where, as one may expect, the energy ordering of the orbitals is different.

It is possible to use the orbitals from both curves as a guess for QED-CASCI calculations. By doing so, two smooth curves can be obtained, where the one obtained from the higher energy solutions is variationally lower with respect to the other curve (see Fig.~S3 in Support Information).
This problem is fixed by QED-CASSCF, as shown in Fig.~\ref{fig:figure3}. In this case, indeed, the variational optimization of the orbitals avoids discontinuities due to the poor quality of the guess, predicting a continuous energy profile in the whole distance range and, as expected, provides variationally lower energies with respect to the QED-CASCI calculations (see Fig.~S4 in Support Information).
This behavior is a demonstration that the presence of the field can, in some cases, complicate the simulation of molecular systems characterized by multireference effects and that the orbital optimization can be important to obtain qualitatively-correct results, especially in the cases where multireference effects become dominant.

\begin{figure}[H]
    \centering
    \includegraphics[width=0.8\linewidth]{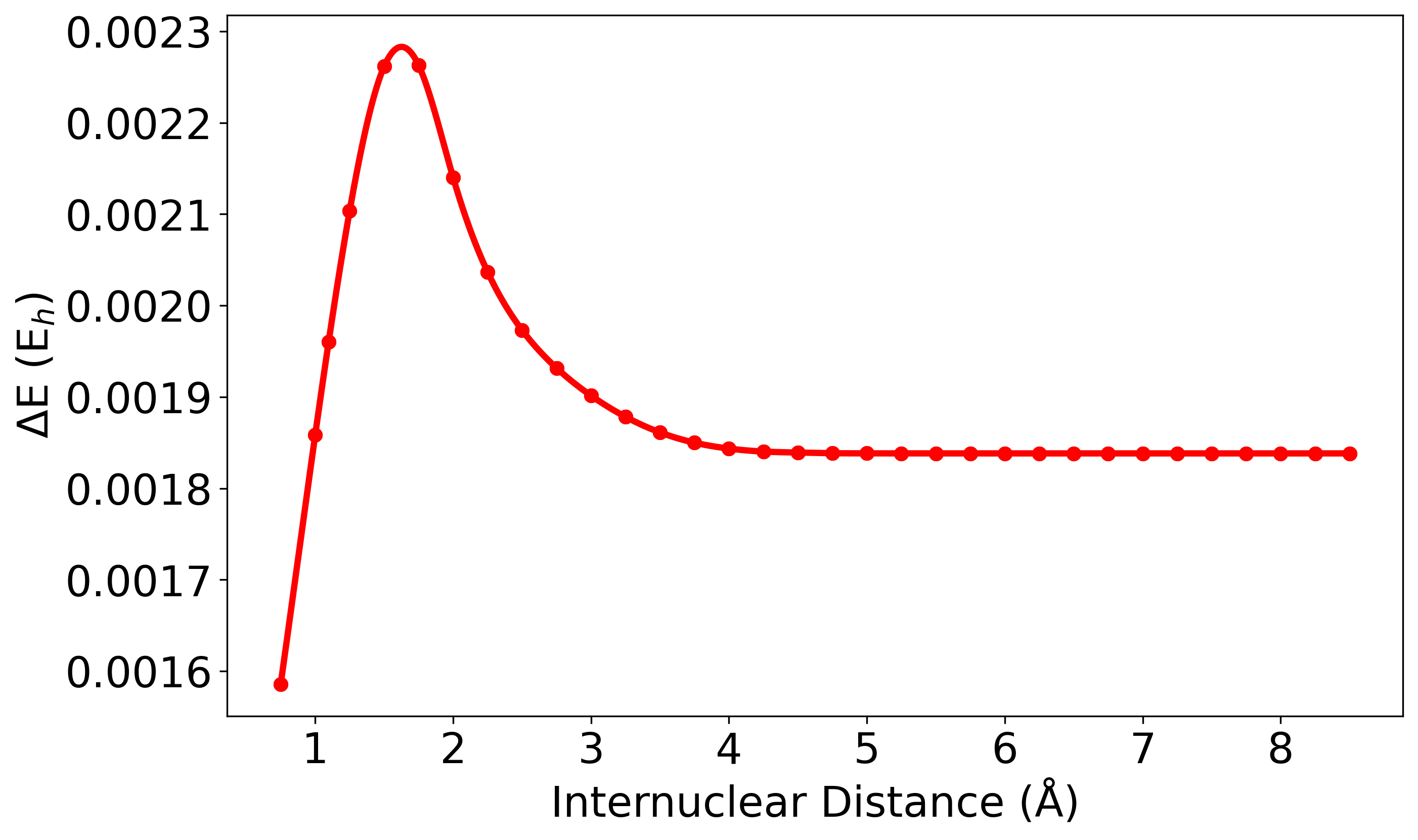}
    \caption{Study of the field-induced effects on the ground state potential energy surfaces computed at the CASSCF level.}\label{fig:figure5}
\end{figure}

In order to estimate the entity of the field-induced effects, we also computed the energy difference between the QED-CASSCF method and the corresponding electronic one (see Fig.~\ref{fig:figure5}) and, as expected, the energy difference goes to a plateau.

Before concluding, it is important to stress that even for very large coupling values (per molecule) like those applied in this test, the field effects on the ground state energy are very small. Much more significant effects could be observed on the excited states energies, but their study goes beyond the scope of this paper and will be analyzed in detail later in a follow-up work.
\subsection{Field-induced effects on the torsion of ethylene}
The description of bond breakings, torsions along conjugated bonds, and conical intersections has always been a challenge for single-reference methods~\cite{mai2020molecular}. CASSCF is still the most commonly employed method to provide a good guess for the description of such phenomena.
In this section we investigate the double bond torsion of an ethylene molecule coupled to an optical cavity.
This molecular system has been largely studied in the literature~\cite{feller2014systematic}. This work does not intend to provide any rigorous quantitative treatment of the chemistry of the system but it only shows the effects induced by the cavity environment.
\begin{figure}[H]
    \centering
    \includegraphics[width=0.6\linewidth]{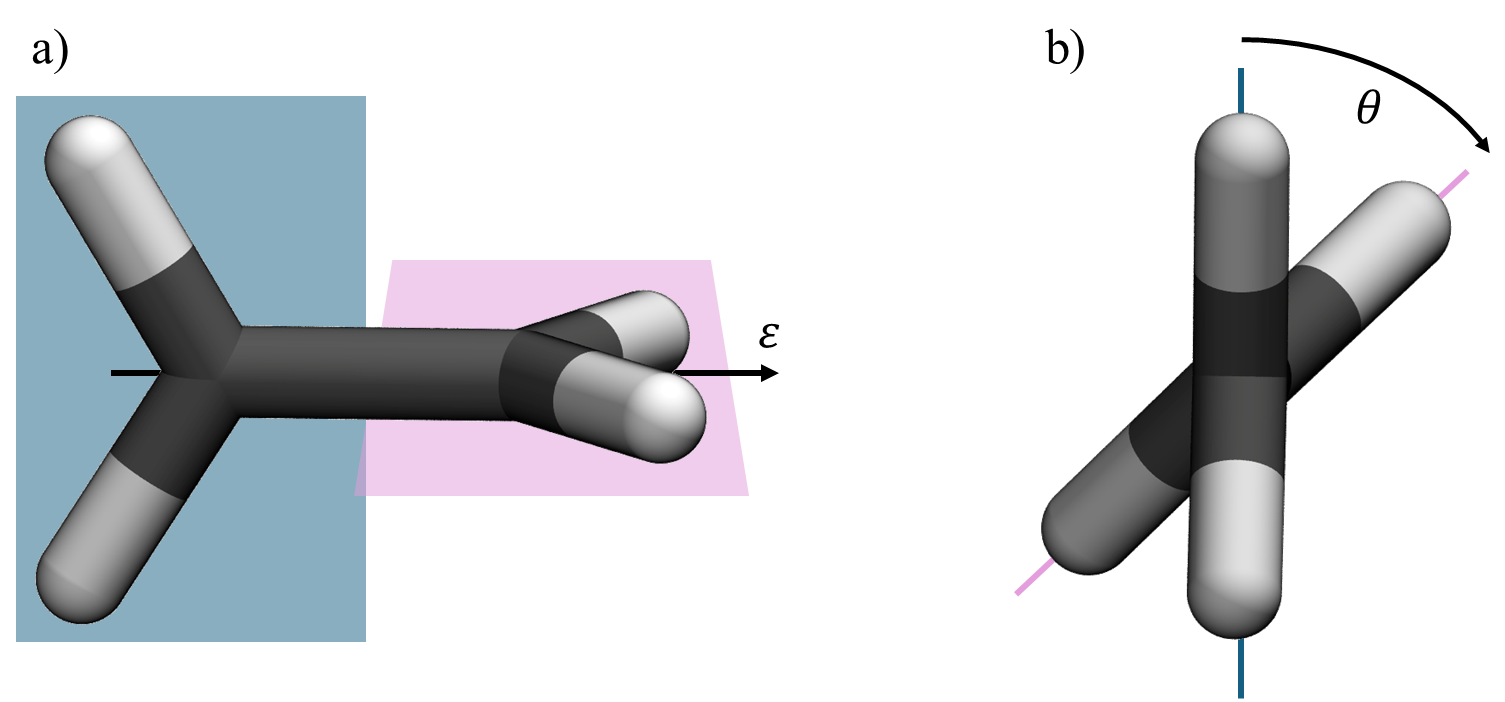}
    \caption{
    \textbf{a.} Twisted ethylene molecule, field polarization $\boldsymbol{\epsilon}$ along the carbon-carbon bond. \textbf{b.} Representation of the dihedral angle, $\theta$.}\label{fig:figure6}
\end{figure}
The torsion of the double bond induces interesting properties in the electronic structure of ethylene, in particular when the dihedral angle approaches 90°. For this particular geometry, the two partially occupied (frontier) orbitals are practically degenerate and have occupation numbers equal to 1. 
This configuration confers a high multiconfigurational character to the system.
In this section we focus our attention on the effects induced by the field on the frontier orbitals energies for ethylene in the $\theta=90$° configuration.
The calculations were performed using a cc-pVDZ basis set and a minimal active space containing the two $\pi$ electrons in the $\pi$ and $\pi^*$ orbitals, CAS(2,2), and only one photon was considered.
In this case, the cavity frequency has been chosen to match the excitation energy of the lowest bright electronic transition computed at the linear response CASSCF level of theory of the $\theta=90$° geometry ($\omega\approx3.30$ eV). The polarization direction has been directed along the main molecular axis ($x$). The orbital energy profiles were recorded by increasing the light-matter coupling $\lambda$ from $0$ to $0.4$ a.u. (see Fig.~\ref{fig:figure7}). When the coupling is increased, the orbital energies remain almost constant until coupling values of about 0.25 a.u. and then change rapidly, reaching a gap of $\sim 4$eV (see Fig.~\ref{fig:figure7}a). This breaking of the orbital degeneracy, with consequent reduction of the multiconfigurational character of the system, reflects also on the orbitals' occupations (see Fig.~\ref{fig:figure7}b). 
Now, one of the frontier orbitals appears occupied with more than one electron, while the other is occupied with small fractions of an electron.
The sudden lifting of the degeneracy of the frontier orbitals at $\lambda=0.29$ a.u. was further investigated.
This behavior can be explained by looking at the energy profiles in Fig.~\ref{fig:figure7}d where a state inversion happens between $\lambda=0.29$ and $0.35$ a.u.
This clearly explains the sudden change in the orbitals' nature, as observed in Fig.~\ref{fig:figure7}c. In this case we can observe a field-induced mixing between the involved molecular orbitals.
Due to the almost degeneracy of the states, convergence to one or the other can be obtained using different starting guesses.
\color{black}
\begin{figure}[H]
    \centering
    \includegraphics[width=1\linewidth]{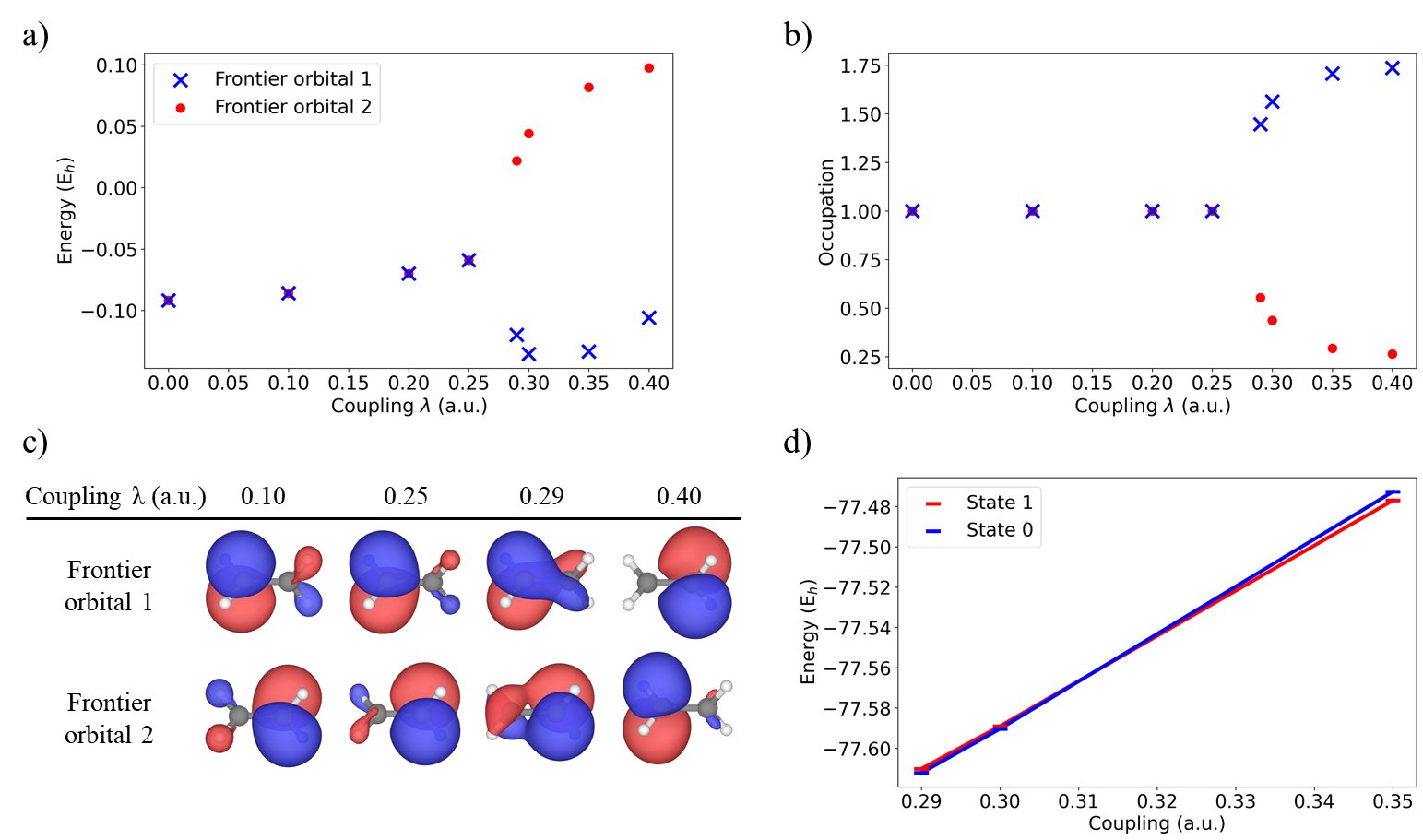}
    \caption{QED-CASSCF frontier orbitals \textbf{a.} energies and \textbf{b.} occupations as a function of the light-matter coupling value. \textbf{c.} Frontier orbitals' shapes for different coupling values. \textbf{d.} Energy profiles of the two states between $\lambda=0.29$ and $0.35$ a.u.}\label{fig:figure7}
\end{figure}
Summarizing, in this paper we observed that in certain conditions the field may induce changes in the multiconfigurational character of a molecular system. The analysis performed in this work has been conducted with light-matter coupling values that are way larger than those currently obtainable in actual experiments. However, if systems with smaller excitation energies are investigated and the method is coupled with strategies to properly account for collective effects, our methodology has the potential to become a reference approach to simulate field-induced effects on the electronic structure of multireference systems.

\subsection{QED-CASSCF origin dependence}
It is now well established (as demonstrated in Ref.~\citenum{haugland2020coupled}) that QED-HF theory has an intrinsic origin dependence if charged systems are investigated.
This origin dependence does not appear on the total energy because of the origin invariance of Hamiltonian~\ref{pf_original} but it is very evident on the molecular orbitals and on their energies. 
This behavior has been addressed in detail by Riso et al. in Ref.~\citenum{riso2022molecular}. In particular, they developed an alternative methodology known as strong coupling quantum electrodynamics Hartree Fock (SC-QED-HF)~\cite{riso2022molecular, el2024toward} with the intent of constructing a consistent molecular orbital theory for molecules in strong coupling conditions. In this framework, the wavefunction is transformed by a unitary transformation, and the orbitals are defined within an electron-photon correlation basis obtained by diagonalizing the operator $(\mathbf{d} \cdot \boldsymbol{\epsilon})$. 
This approach provides fully origin-invariant molecular orbitals and is able to recover part of electron-photon correlation energy. Therefore, SC-QED-HF is a better fit for the investigation of cavity-induced effects and provides a physically meaningful guess for performing post-HF calculations on charged systems.
Recently, the strong coupling formalism has also been extended to second-order M{\o}ller-Plesset theory and response theory~\cite{moutaoukal2025strong, castagnola2025strong}.

Since also CASSCF involves an optimization of the molecular orbitals, it makes sense to check whether the origin invariance problem also affect this method when charged systems are investigated.

\begin{figure}[H]
    \centering
    \includegraphics[width=1.0\linewidth]{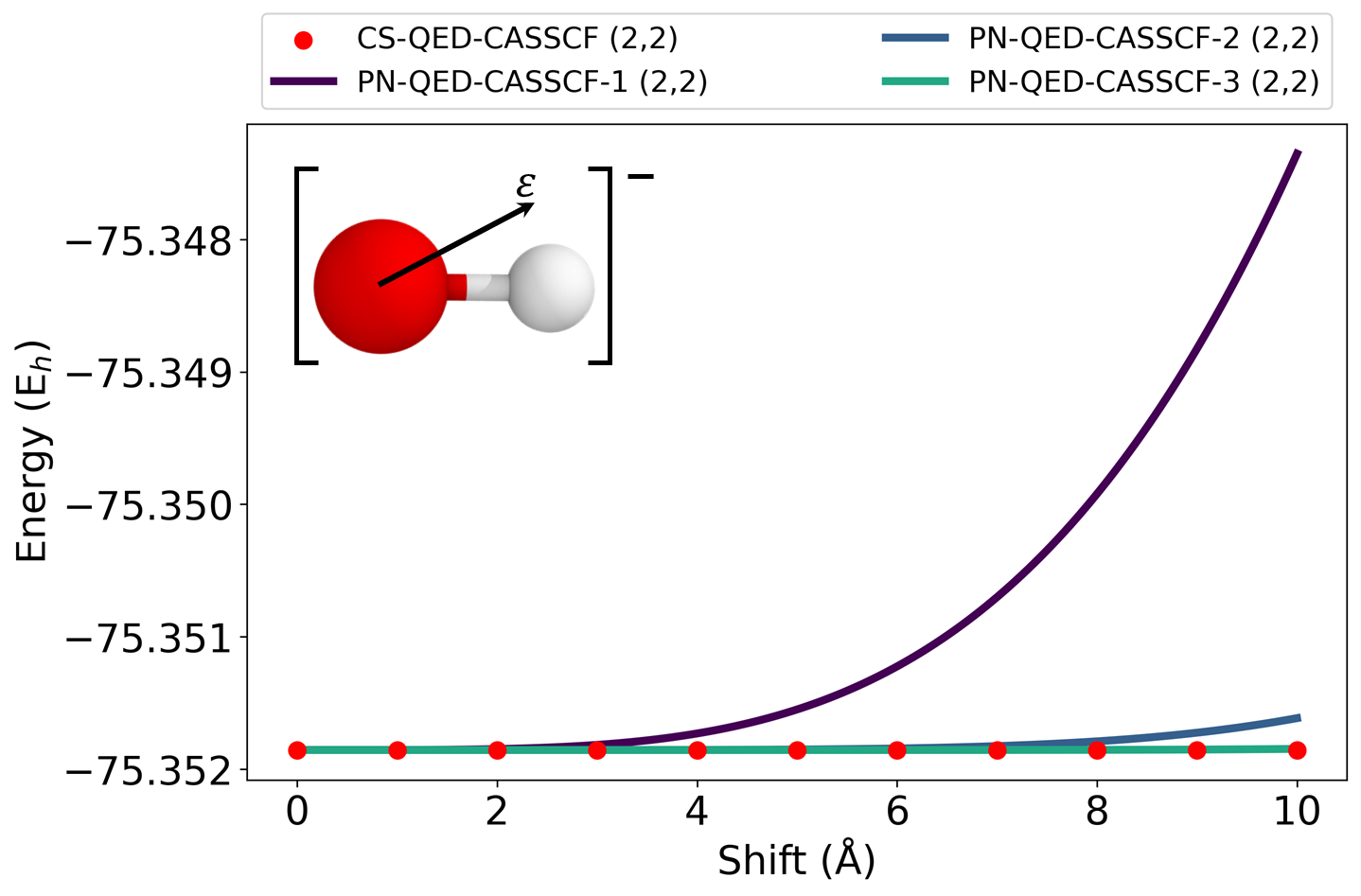}
    \caption{Profile of the QED-CASSCF energy computed with the PN-Hamiltonian and compared with the CS-transformed Hamiltonian for the OH$^{-}$ ion.}
    \label{fig:figure8}
\end{figure}

In this section we address this point in detail. In particular, we performed a simple test case on the hydroxide ion. Here, CAS(2,2) calculations were performed while shifting the molecule along the $z$ axis (the origin of the reference system has been centered on the oxygen atom). The cavity frequency was set, in this case, to $\omega=0.5$ eV, the light-matter coupling to $\lambda=0.03$ a.u., and the polarization vector is oriented along the direction $\mathbf{\epsilon}=(\frac{1}{\sqrt{3}},\frac{1}{\sqrt{3}},\frac{1}{\sqrt{3}})$.
We compared the energy of the coherent-state (CS) transformed Hamiltonian (Eq.~\ref{pf_original}) to its untransformed version (obtained by setting $\braket{\hat{d}}=0$),
often known as photon-number (PN) Hamiltonian.
Results are presented in Fig.~\ref{fig:figure8}.
In this plot we see that while the total energy computed with the CS-Hamiltonian is origin independent, the results obtained by PN-QED-CASSCF have, as expected, an explicit dependence on the choice of the reference system.
This dependence disappears if a large number of photonic states is included into the treatment. However, considering that increasing the dimension of the photonic Fock space results in a higher computational costs, the coherent-state transformation offers a more cost-effective and reliable strategy for the treatment of these systems. 

It is important to point out that the origin dependence should still be present in the molecular orbitals. This can be expected by the explicit origin dependence of the QED-CASSCF Fock matrix of Eq.~\ref{bilinear_orbital_gradient}. This problem can be eventually fixed by applying a unitary transformation similar to the one used in Ref.~\citenum{riso2022molecular} to the CASSCF procedure. However, the application of this transformation to QED-CASSCF goes, for the moment, beyond the scope of this paper.
\color{black}
\section{Conclusions\label{section:5}}
In this work we have presented an extension of the complete active space self-consistent field method to account for the effect of quantized electromagnetic fields in optical cavities (QED-CASSCF). The implementation of the method was performed according to a restricted-step second-order optimization method. This strategy allows us to overcome some of the convergence issues that characterize the CASSCF wavefunction optimization~\cite{nottoli2021second, helmich2022trust}.  

The developed method was tested on small multireference molecules where the inclusion of static correlation is crucial to provide at least a qualitative description of the system.
The comparison between QED-CASSCF and QED-CASCI for the dissociation curve of the nitrogen molecule proved the importance of the orbital optimization to avoid unphysical behaviors arising from possible poor quality guesses.

We also investigated the effects induced by the cavity field on the ethylene torsion process. Our calculations show that the presence of the electromagnetic field can remove the degeneracy between the 
$\pi$ and $\pi^*$ orbitals reducing, in specific coupling conditions, the multireference character of the system.

Finally, we addressed the origin dependence of the method in the case of charged systems. 
In particular, the simple case study of the OH$^-$ ion showed that, when the CS-Hamiltonian is used, the total energy is invariant with respect to the choice of origin. 
In contrast, the photon number PN-Hamiltonian exhibits origin dependence, although the energy converges toward the origin-independent result as the size of the photonic Fock space increases. 
However, the origin dependence of the molecular orbitals persists even when employing the CS-transformed Hamiltonian.
This issue will be solved in a future paper where the Strong Coupling transformation proposed in Ref.~\citenum{riso2022molecular} will be applied to remove the origin-dependence of QED-CASSCF also for charged systems.

It is important to remind that, the results obtained in this paper, due to the very large coupling values we used, are not directly comparable with experimental results.
However, the aim of the paper, was not much to provide accurate simulations of molecular systems in optical cavities, but more to propose a new methodology able to investigate multireference systems coupled to electromagnetic fields. In the near 
future this methodology will be coupled with strategies to include collective effects that will allow for more realistic simulations~\cite{castagnola2024realistic, koessler2025polariton, horak2025analytic}.

With this work we move one step further towards an accurate description of multiconfigurational molecular systems in optical cavities. This paves the way for the investigation of chemically/photochemically-relevant processes that until this day have been investigated through model Hamiltonians or other approximated methods~\cite{galego2015cavity, bennett2016novel, rana2023simulating}.
\begin{acknowledgement}
We thank Sarai Dery Folkestad and Jonathan Foley for insightful discussions.
R.A and E.R. acknowledge funding from the European Research Council (ERC) under the European Union’s Horizon Europe Research and Innovation Programme (Grant n. ERC-StG-2021-101040197 - QED-SPIN).
H.K. acknowledges funding from the European Research Council (ERC) under the European Union’s Horizon 2020 Research and Innovation Programme (grant agreement No. 101020016). 
\end{acknowledgement}
\begin{suppinfo}
The SI contains the plot of the QED-CASSCF potential energy curve for the nitrogen molecule computed with an increasing number of photons, the comparison between the CASCI and QED-CASCI curves obtained from independent guesses, and the QED-CASCI curves computed using the high and low energy QED-HF guesses.

\subsection{Code availability}
The $e^\mathcal{T}$ program used to perform the calculations shown in this work is available from the corresponding author upon reasonable request.
Examples of the input files used to run the calculations are available at the following Zenodo link: \href{https://doi.org/10.5281/zenodo.15118476}{https://doi.org/10.5281/zenodo.15118476}.

\end{suppinfo}


\begin{mcitethebibliography}{113}
\providecommand*\natexlab[1]{#1}
\providecommand*\mciteSetBstSublistMode[1]{}
\providecommand*\mciteSetBstMaxWidthForm[2]{}
\providecommand*\mciteBstWouldAddEndPuncttrue
  {\def\EndOfBibitem{\unskip.}}
\providecommand*\mciteBstWouldAddEndPunctfalse
  {\let\EndOfBibitem\relax}
\providecommand*\mciteSetBstMidEndSepPunct[3]{}
\providecommand*\mciteSetBstSublistLabelBeginEnd[3]{}
\providecommand*\EndOfBibitem{}
\mciteSetBstSublistMode{f}
\mciteSetBstMaxWidthForm{subitem}{(\alph{mcitesubitemcount})}
\mciteSetBstSublistLabelBeginEnd
  {\mcitemaxwidthsubitemform\space}
  {\relax}
  {\relax}

\bibitem[Ebbesen(2016)]{ebbesen2016hybrid}
Ebbesen,~T.~W. {Hybrid light--matter states in a molecular and material science perspective}. \emph{Acc. Chem. Res.} \textbf{2016}, \emph{49}, 2403--2412\relax
\mciteBstWouldAddEndPuncttrue
\mciteSetBstMidEndSepPunct{\mcitedefaultmidpunct}
{\mcitedefaultendpunct}{\mcitedefaultseppunct}\relax
\EndOfBibitem
\bibitem[Herrera and Spano(2016)Herrera, and Spano]{herrera2016cavity}
Herrera,~F.; Spano,~F.~C. {Cavity-controlled chemistry in molecular ensembles}. \emph{Phys. Rev. Lett.} \textbf{2016}, \emph{116}, 238301\relax
\mciteBstWouldAddEndPuncttrue
\mciteSetBstMidEndSepPunct{\mcitedefaultmidpunct}
{\mcitedefaultendpunct}{\mcitedefaultseppunct}\relax
\EndOfBibitem
\bibitem[Galego \latin{et~al.}(2017)Galego, Garcia-Vidal, and Feist]{galego2017many}
Galego,~J.; Garcia-Vidal,~F.~J.; Feist,~J. {Many-molecule reaction triggered by a single photon in polaritonic chemistry}. \emph{Phys. Rev. Lett.} \textbf{2017}, \emph{119}, 136001\relax
\mciteBstWouldAddEndPuncttrue
\mciteSetBstMidEndSepPunct{\mcitedefaultmidpunct}
{\mcitedefaultendpunct}{\mcitedefaultseppunct}\relax
\EndOfBibitem
\bibitem[Mart{\'\i}nez-Mart{\'\i}nez \latin{et~al.}(2018)Mart{\'\i}nez-Mart{\'\i}nez, Du, Ribeiro, K{\'e}na-Cohen, and Yuen-Zhou]{martinez2018polariton}
Mart{\'\i}nez-Mart{\'\i}nez,~L.~A.; Du,~M.; Ribeiro,~R.~F.; K{\'e}na-Cohen,~S.; Yuen-Zhou,~J. {Polariton-assisted singlet fission in acene aggregates}. \emph{J. Phys. Chem. Lett.} \textbf{2018}, \emph{9}, 1951--1957\relax
\mciteBstWouldAddEndPuncttrue
\mciteSetBstMidEndSepPunct{\mcitedefaultmidpunct}
{\mcitedefaultendpunct}{\mcitedefaultseppunct}\relax
\EndOfBibitem
\bibitem[Mandal and Huo(2019)Mandal, and Huo]{mandal2019investigating}
Mandal,~A.; Huo,~P. {Investigating new reactivities enabled by polariton photochemistry}. \emph{J. Phys. Chem. Lett.} \textbf{2019}, \emph{10}, 5519--5529\relax
\mciteBstWouldAddEndPuncttrue
\mciteSetBstMidEndSepPunct{\mcitedefaultmidpunct}
{\mcitedefaultendpunct}{\mcitedefaultseppunct}\relax
\EndOfBibitem
\bibitem[Wang \latin{et~al.}(2020)Wang, Scholes, and Hsu]{wang2020coherent}
Wang,~S.; Scholes,~G.~D.; Hsu,~L.-Y. {Coherent-to-incoherent transition of molecular fluorescence controlled by surface plasmon polaritons}. \emph{J. Phys. Chem. Lett.} \textbf{2020}, \emph{11}, 5948--5955\relax
\mciteBstWouldAddEndPuncttrue
\mciteSetBstMidEndSepPunct{\mcitedefaultmidpunct}
{\mcitedefaultendpunct}{\mcitedefaultseppunct}\relax
\EndOfBibitem
\bibitem[DelPo \latin{et~al.}(2020)DelPo, Kudisch, Park, Khan, Fassioli, Fausti, Rand, and Scholes]{delpo2020polariton}
DelPo,~C.~A.; Kudisch,~B.; Park,~K.~H.; Khan,~S.-U.-Z.; Fassioli,~F.; Fausti,~D.; Rand,~B.~P.; Scholes,~G.~D. {Polariton transitions in femtosecond transient absorption studies of ultrastrong light--molecule coupling}. \emph{J. Phys. Chem. Lett.} \textbf{2020}, \emph{11}, 2667--2674\relax
\mciteBstWouldAddEndPuncttrue
\mciteSetBstMidEndSepPunct{\mcitedefaultmidpunct}
{\mcitedefaultendpunct}{\mcitedefaultseppunct}\relax
\EndOfBibitem
\bibitem[Polak \latin{et~al.}(2020)Polak, Jayaprakash, Lyons, Mart{\'\i}nez-Mart{\'\i}nez, Leventis, Fallon, Coulthard, Bossanyi, Georgiou, Petty, \latin{et~al.} others]{polak2020manipulating}
Polak,~D.; Jayaprakash,~R.; Lyons,~T.~P.; Mart{\'\i}nez-Mart{\'\i}nez,~L.~{\'A}.; Leventis,~A.; Fallon,~K.~J.; Coulthard,~H.; Bossanyi,~D.~G.; Georgiou,~K.; Petty,~A.~J.; others {Manipulating molecules with strong coupling: harvesting triplet excitons in organic exciton microcavities}. \emph{Chem. Sci.} \textbf{2020}, \emph{11}, 343--354\relax
\mciteBstWouldAddEndPuncttrue
\mciteSetBstMidEndSepPunct{\mcitedefaultmidpunct}
{\mcitedefaultendpunct}{\mcitedefaultseppunct}\relax
\EndOfBibitem
\bibitem[Thompson \latin{et~al.}(1992)Thompson, Rempe, and Kimble]{thompson1992observation}
Thompson,~R.; Rempe,~G.; Kimble,~H. {Observation of normal-mode splitting for an atom in an optical cavity}. \emph{Phys. Rev. Lett.} \textbf{1992}, \emph{68}, 1132\relax
\mciteBstWouldAddEndPuncttrue
\mciteSetBstMidEndSepPunct{\mcitedefaultmidpunct}
{\mcitedefaultendpunct}{\mcitedefaultseppunct}\relax
\EndOfBibitem
\bibitem[Garcia-Vidal \latin{et~al.}(2021)Garcia-Vidal, Ciuti, and Ebbesen]{garcia2021manipulating}
Garcia-Vidal,~F.~J.; Ciuti,~C.; Ebbesen,~T.~W. {Manipulating matter by strong coupling to vacuum fields}. \emph{Science} \textbf{2021}, \emph{373}, eabd0336\relax
\mciteBstWouldAddEndPuncttrue
\mciteSetBstMidEndSepPunct{\mcitedefaultmidpunct}
{\mcitedefaultendpunct}{\mcitedefaultseppunct}\relax
\EndOfBibitem
\bibitem[Baumberg \latin{et~al.}(2019)Baumberg, Aizpurua, Mikkelsen, and Smith]{baumberg2019extreme}
Baumberg,~J.~J.; Aizpurua,~J.; Mikkelsen,~M.~H.; Smith,~D.~R. {Extreme nanophotonics from ultrathin metallic gaps}. \emph{Nat. Mat.} \textbf{2019}, \emph{18}, 668--678\relax
\mciteBstWouldAddEndPuncttrue
\mciteSetBstMidEndSepPunct{\mcitedefaultmidpunct}
{\mcitedefaultendpunct}{\mcitedefaultseppunct}\relax
\EndOfBibitem
\bibitem[Baranov \latin{et~al.}(2020)Baranov, Munkhbat, Zhukova, Bisht, Canales, Rousseaux, Johansson, Antosiewicz, and Shegai]{baranov2020ultrastrong}
Baranov,~D.~G.; Munkhbat,~B.; Zhukova,~E.; Bisht,~A.; Canales,~A.; Rousseaux,~B.; Johansson,~G.; Antosiewicz,~T.~J.; Shegai,~T. {Ultrastrong coupling between nanoparticle plasmons and cavity photons at ambient conditions}. \emph{Nat. Comm.} \textbf{2020}, \emph{11}, 2715\relax
\mciteBstWouldAddEndPuncttrue
\mciteSetBstMidEndSepPunct{\mcitedefaultmidpunct}
{\mcitedefaultendpunct}{\mcitedefaultseppunct}\relax
\EndOfBibitem
\bibitem[Teufel \latin{et~al.}(2011)Teufel, Li, Allman, Cicak, Sirois, Whittaker, and Simmonds]{teufel2011circuit}
Teufel,~J.~D.; Li,~D.; Allman,~M.~S.; Cicak,~K.; Sirois,~A.; Whittaker,~J.~D.; Simmonds,~R. {Circuit cavity electromechanics in the strong-coupling regime}. \emph{Nature} \textbf{2011}, \emph{471}, 204--208\relax
\mciteBstWouldAddEndPuncttrue
\mciteSetBstMidEndSepPunct{\mcitedefaultmidpunct}
{\mcitedefaultendpunct}{\mcitedefaultseppunct}\relax
\EndOfBibitem
\bibitem[Haroche \latin{et~al.}(2020)Haroche, Brune, and Raimond]{haroche2020cavity}
Haroche,~S.; Brune,~M.; Raimond,~J. {From cavity to circuit quantum electrodynamics}. \emph{Nat. Phys.} \textbf{2020}, \emph{16}, 243--246\relax
\mciteBstWouldAddEndPuncttrue
\mciteSetBstMidEndSepPunct{\mcitedefaultmidpunct}
{\mcitedefaultendpunct}{\mcitedefaultseppunct}\relax
\EndOfBibitem
\bibitem[Wallraff \latin{et~al.}(2004)Wallraff, Schuster, Blais, Frunzio, Huang, Majer, Kumar, Girvin, and Schoelkopf]{wallraff2004strong}
Wallraff,~A.; Schuster,~D.~I.; Blais,~A.; Frunzio,~L.; Huang,~R.-S.; Majer,~J.; Kumar,~S.; Girvin,~S.~M.; Schoelkopf,~R.~J. {Strong coupling of a single photon to a superconducting qubit using circuit quantum electrodynamics}. \emph{Nature} \textbf{2004}, \emph{431}, 162--167\relax
\mciteBstWouldAddEndPuncttrue
\mciteSetBstMidEndSepPunct{\mcitedefaultmidpunct}
{\mcitedefaultendpunct}{\mcitedefaultseppunct}\relax
\EndOfBibitem
\bibitem[Sun \latin{et~al.}(2021)Sun, Huang, Zheng, Xu, Ke, Zhan, Chen, and Deng]{sun2021polariton}
Sun,~F.; Huang,~W.; Zheng,~Z.; Xu,~N.; Ke,~Y.; Zhan,~R.; Chen,~H.; Deng,~S. {Polariton waveguide modes in two-dimensional van der Waals crystals: an analytical model and correlative nano-imaging}. \emph{Nanoscale} \textbf{2021}, \emph{13}, 4845--4854\relax
\mciteBstWouldAddEndPuncttrue
\mciteSetBstMidEndSepPunct{\mcitedefaultmidpunct}
{\mcitedefaultendpunct}{\mcitedefaultseppunct}\relax
\EndOfBibitem
\bibitem[Downing \latin{et~al.}(2019)Downing, Sturges, Weick, Stobi{\'n}ska, and Mart{\'\i}n-Moreno]{downing2019topological}
Downing,~C.; Sturges,~T.; Weick,~G.; Stobi{\'n}ska,~M.; Mart{\'\i}n-Moreno,~L. {Topological phases of polaritons in a cavity waveguide}. \emph{Phys. Rev. Lett.} \textbf{2019}, \emph{123}, 217401\relax
\mciteBstWouldAddEndPuncttrue
\mciteSetBstMidEndSepPunct{\mcitedefaultmidpunct}
{\mcitedefaultendpunct}{\mcitedefaultseppunct}\relax
\EndOfBibitem
\bibitem[Kondratyev \latin{et~al.}(2023)Kondratyev, Permyakov, Ivanova, Iorsh, Krizhanovskii, Skolnick, Kravtsov, and Samusev]{kondratyev2023probing}
Kondratyev,~V.~I.; Permyakov,~D.~V.; Ivanova,~T.~V.; Iorsh,~I.~V.; Krizhanovskii,~D.~N.; Skolnick,~M.~S.; Kravtsov,~V.; Samusev,~A.~K. {Probing and control of guided exciton--polaritons in a 2D semiconductor-integrated slab waveguide}. \emph{Nano Lett.} \textbf{2023}, \emph{23}, 7876--7882\relax
\mciteBstWouldAddEndPuncttrue
\mciteSetBstMidEndSepPunct{\mcitedefaultmidpunct}
{\mcitedefaultendpunct}{\mcitedefaultseppunct}\relax
\EndOfBibitem
\bibitem[Miller \latin{et~al.}(2005)Miller, Northup, Birnbaum, Boca, Boozer, and Kimble]{miller2005trapped}
Miller,~R.; Northup,~T.; Birnbaum,~K.; Boca,~A.; Boozer,~A.; Kimble,~H. {Trapped atoms in cavity QED: coupling quantized light and matter}. \emph{J. Phys. B} \textbf{2005}, \emph{38}, S551\relax
\mciteBstWouldAddEndPuncttrue
\mciteSetBstMidEndSepPunct{\mcitedefaultmidpunct}
{\mcitedefaultendpunct}{\mcitedefaultseppunct}\relax
\EndOfBibitem
\bibitem[Flick \latin{et~al.}(2017)Flick, Ruggenthaler, Appel, and Rubio]{flick2017atoms}
Flick,~J.; Ruggenthaler,~M.; Appel,~H.; Rubio,~A. {Atoms and molecules in cavities, from weak to strong coupling in quantum-electrodynamics (QED) chemistry}. \emph{PNAS} \textbf{2017}, \emph{114}, 3026--3034\relax
\mciteBstWouldAddEndPuncttrue
\mciteSetBstMidEndSepPunct{\mcitedefaultmidpunct}
{\mcitedefaultendpunct}{\mcitedefaultseppunct}\relax
\EndOfBibitem
\bibitem[Feist \latin{et~al.}(2018)Feist, Galego, and Garcia-Vidal]{feist2018polaritonic}
Feist,~J.; Galego,~J.; Garcia-Vidal,~F.~J. {Polaritonic chemistry with organic molecules}. \emph{ACS Photonics} \textbf{2018}, \emph{5}, 205--216\relax
\mciteBstWouldAddEndPuncttrue
\mciteSetBstMidEndSepPunct{\mcitedefaultmidpunct}
{\mcitedefaultendpunct}{\mcitedefaultseppunct}\relax
\EndOfBibitem
\bibitem[Hertzog \latin{et~al.}(2019)Hertzog, Wang, Mony, and B{\"o}rjesson]{hertzog2019strong}
Hertzog,~M.; Wang,~M.; Mony,~J.; B{\"o}rjesson,~K. {Strong light--matter interactions: a new direction within chemistry}. \emph{Chem. Soc. Rev.} \textbf{2019}, \emph{48}, 937--961\relax
\mciteBstWouldAddEndPuncttrue
\mciteSetBstMidEndSepPunct{\mcitedefaultmidpunct}
{\mcitedefaultendpunct}{\mcitedefaultseppunct}\relax
\EndOfBibitem
\bibitem[Herrera and Spano(2017)Herrera, and Spano]{herrera2017absorption}
Herrera,~F.; Spano,~F.~C. {Absorption and photoluminescence in organic cavity QED}. \emph{Phys. Rev. A} \textbf{2017}, \emph{95}, 053867\relax
\mciteBstWouldAddEndPuncttrue
\mciteSetBstMidEndSepPunct{\mcitedefaultmidpunct}
{\mcitedefaultendpunct}{\mcitedefaultseppunct}\relax
\EndOfBibitem
\bibitem[Baranov \latin{et~al.}(2020)Baranov, Munkhbat, L{\"a}nk, Verre, K{\"a}ll, and Shegai]{baranov2020circular}
Baranov,~D.~G.; Munkhbat,~B.; L{\"a}nk,~N.~O.; Verre,~R.; K{\"a}ll,~M.; Shegai,~T. {Circular dichroism mode splitting and bounds to its enhancement with cavity-plasmon-polaritons}. \emph{Nanophotonics} \textbf{2020}, \emph{9}, 283--293\relax
\mciteBstWouldAddEndPuncttrue
\mciteSetBstMidEndSepPunct{\mcitedefaultmidpunct}
{\mcitedefaultendpunct}{\mcitedefaultseppunct}\relax
\EndOfBibitem
\bibitem[Shalabney \latin{et~al.}(2015)Shalabney, George, Hiura, Hutchison, Genet, Hellwig, and Ebbesen]{shalabney2015enhanced}
Shalabney,~A.; George,~J.; Hiura,~H.; Hutchison,~J.~A.; Genet,~C.; Hellwig,~P.; Ebbesen,~T.~W. {Enhanced raman scattering from vibro-polariton hybrid states}. \emph{Angew Chem. Int. Ed. Engl.} \textbf{2015}, \emph{54}, 7971--7975\relax
\mciteBstWouldAddEndPuncttrue
\mciteSetBstMidEndSepPunct{\mcitedefaultmidpunct}
{\mcitedefaultendpunct}{\mcitedefaultseppunct}\relax
\EndOfBibitem
\bibitem[del Pino \latin{et~al.}(2015)del Pino, Feist, and Garcia-Vidal]{del2015signatures}
del Pino,~J.; Feist,~J.; Garcia-Vidal,~F. {Signatures of vibrational strong coupling in Raman scattering}. \emph{J. Phys. Chem. C} \textbf{2015}, \emph{119}, 29132--29137\relax
\mciteBstWouldAddEndPuncttrue
\mciteSetBstMidEndSepPunct{\mcitedefaultmidpunct}
{\mcitedefaultendpunct}{\mcitedefaultseppunct}\relax
\EndOfBibitem
\bibitem[Skolnick \latin{et~al.}(1998)Skolnick, Fisher, and Whittaker]{skolnick1998strong}
Skolnick,~M.; Fisher,~T.; Whittaker,~D. {Strong coupling phenomena in quantum microcavity structures}. \emph{Semicond. Sci. Technol.} \textbf{1998}, \emph{13}, 645\relax
\mciteBstWouldAddEndPuncttrue
\mciteSetBstMidEndSepPunct{\mcitedefaultmidpunct}
{\mcitedefaultendpunct}{\mcitedefaultseppunct}\relax
\EndOfBibitem
\bibitem[Kasprzak \latin{et~al.}(2006)Kasprzak, Richard, Kundermann, Baas, Jeambrun, Keeling, Marchetti, Szyma{\'n}ska, Andr{\'e}, Staehli, \latin{et~al.} others]{kasprzak2006bose}
Kasprzak,~J.; Richard,~M.; Kundermann,~S.; Baas,~A.; Jeambrun,~P.; Keeling,~J. M.~J.; Marchetti,~F.; Szyma{\'n}ska,~M.; Andr{\'e},~R.; Staehli,~J.~a.; others {Bose--Einstein condensation of exciton polaritons}. \emph{Nature} \textbf{2006}, \emph{443}, 409--414\relax
\mciteBstWouldAddEndPuncttrue
\mciteSetBstMidEndSepPunct{\mcitedefaultmidpunct}
{\mcitedefaultendpunct}{\mcitedefaultseppunct}\relax
\EndOfBibitem
\bibitem[Byrnes \latin{et~al.}(2014)Byrnes, Kim, and Yamamoto]{byrnes2014exciton}
Byrnes,~T.; Kim,~N.~Y.; Yamamoto,~Y. {Exciton--polariton condensates}. \emph{Nat. Phys.} \textbf{2014}, \emph{10}, 803--813\relax
\mciteBstWouldAddEndPuncttrue
\mciteSetBstMidEndSepPunct{\mcitedefaultmidpunct}
{\mcitedefaultendpunct}{\mcitedefaultseppunct}\relax
\EndOfBibitem
\bibitem[Orgiu \latin{et~al.}(2015)Orgiu, George, Hutchison, Devaux, Dayen, Doudin, Stellacci, Genet, Schachenmayer, Genes, \latin{et~al.} others]{orgiu2015conductivity}
Orgiu,~E.; George,~J.; Hutchison,~J.~A.; Devaux,~E.; Dayen,~J.-F.; Doudin,~B.; Stellacci,~F.; Genet,~C.; Schachenmayer,~J.; Genes,~C.; others {Conductivity in organic semiconductors hybridized with the vacuum field}. \emph{Nat. Mat.} \textbf{2015}, \emph{14}, 1123--1129\relax
\mciteBstWouldAddEndPuncttrue
\mciteSetBstMidEndSepPunct{\mcitedefaultmidpunct}
{\mcitedefaultendpunct}{\mcitedefaultseppunct}\relax
\EndOfBibitem
\bibitem[Bienfait \latin{et~al.}(2016)Bienfait, Pla, Kubo, Zhou, Stern, Lo, Weis, Schenkel, Vion, Esteve, \latin{et~al.} others]{bienfait2016controlling}
Bienfait,~A.; Pla,~J.; Kubo,~Y.; Zhou,~X.; Stern,~M.; Lo,~C.; Weis,~C.; Schenkel,~T.; Vion,~D.; Esteve,~D.; others {Controlling spin relaxation with a cavity}. \emph{Nature} \textbf{2016}, \emph{531}, 74--77\relax
\mciteBstWouldAddEndPuncttrue
\mciteSetBstMidEndSepPunct{\mcitedefaultmidpunct}
{\mcitedefaultendpunct}{\mcitedefaultseppunct}\relax
\EndOfBibitem
\bibitem[Bonizzoni \latin{et~al.}(2017)Bonizzoni, Ghirri, Atzori, Sorace, Sessoli, and Affronte]{bonizzoni2017coherent}
Bonizzoni,~C.; Ghirri,~A.; Atzori,~M.; Sorace,~L.; Sessoli,~R.; Affronte,~M. {Coherent coupling between Vanadyl Phthalocyanine spin ensemble and microwave photons: towards integration of molecular spin qubits into quantum circuits}. \emph{Sci. Rep.} \textbf{2017}, \emph{7}, 13096\relax
\mciteBstWouldAddEndPuncttrue
\mciteSetBstMidEndSepPunct{\mcitedefaultmidpunct}
{\mcitedefaultendpunct}{\mcitedefaultseppunct}\relax
\EndOfBibitem
\bibitem[Bonizzoni \latin{et~al.}(2018)Bonizzoni, Ghirri, and Affronte]{bonizzoni2018coherent}
Bonizzoni,~C.; Ghirri,~A.; Affronte,~M. {Coherent coupling of molecular spins with microwave photons in planar superconducting resonators}. \emph{Adv. Phys.: X} \textbf{2018}, \emph{3}, 1435305\relax
\mciteBstWouldAddEndPuncttrue
\mciteSetBstMidEndSepPunct{\mcitedefaultmidpunct}
{\mcitedefaultendpunct}{\mcitedefaultseppunct}\relax
\EndOfBibitem
\bibitem[Thomas \latin{et~al.}(2016)Thomas, George, Shalabney, Dryzhakov, Varma, Moran, Chervy, Zhong, Devaux, Genet, \latin{et~al.} others]{ebbesen2016ground}
Thomas,~A.; George,~J.; Shalabney,~A.; Dryzhakov,~M.; Varma,~S.~J.; Moran,~J.; Chervy,~T.; Zhong,~X.; Devaux,~E.; Genet,~C.; others {Ground-state chemical reactivity under vibrational coupling to the vacuum electromagnetic field}. \emph{Angew. Chem.} \textbf{2016}, \emph{128}, 11634--11638\relax
\mciteBstWouldAddEndPuncttrue
\mciteSetBstMidEndSepPunct{\mcitedefaultmidpunct}
{\mcitedefaultendpunct}{\mcitedefaultseppunct}\relax
\EndOfBibitem
\bibitem[Lather \latin{et~al.}(2019)Lather, Bhatt, Thomas, Ebbesen, and George]{ebbesen2019cavity}
Lather,~J.; Bhatt,~P.; Thomas,~A.; Ebbesen,~T.~W.; George,~J. {Cavity catalysis by cooperative vibrational strong coupling of reactant and solvent molecules}. \emph{Angew Chem. Int. Ed. Engl.} \textbf{2019}, \emph{58}, 10635--10638\relax
\mciteBstWouldAddEndPuncttrue
\mciteSetBstMidEndSepPunct{\mcitedefaultmidpunct}
{\mcitedefaultendpunct}{\mcitedefaultseppunct}\relax
\EndOfBibitem
\bibitem[Thomas \latin{et~al.}(2019)Thomas, Lethuillier-Karl, Nagarajan, Vergauwe, George, Chervy, Shalabney, Devaux, Genet, Moran, and Ebbesen]{thomas2019tilting}
Thomas,~A.; Lethuillier-Karl,~L.; Nagarajan,~K.; Vergauwe,~R.~M.; George,~J.; Chervy,~T.; Shalabney,~A.; Devaux,~E.; Genet,~C.; Moran,~J.; Ebbesen,~T.~W. {Tilting a ground-state reactivity landscape by vibrational strong coupling}. \emph{Science} \textbf{2019}, \emph{363}, 615--619\relax
\mciteBstWouldAddEndPuncttrue
\mciteSetBstMidEndSepPunct{\mcitedefaultmidpunct}
{\mcitedefaultendpunct}{\mcitedefaultseppunct}\relax
\EndOfBibitem
\bibitem[Sau \latin{et~al.}(2021)Sau, Nagarajan, Patrahau, Lethuillier-Karl, Vergauwe, Thomas, Moran, Genet, and Ebbesen]{sau2021modifying}
Sau,~A.; Nagarajan,~K.; Patrahau,~B.; Lethuillier-Karl,~L.; Vergauwe,~R.~M.; Thomas,~A.; Moran,~J.; Genet,~C.; Ebbesen,~T.~W. {Modifying Woodward--Hoffmann stereoselectivity under vibrational strong coupling}. \emph{Angew Chem. Int. Ed. Engl.} \textbf{2021}, \emph{60}, 5712--5717\relax
\mciteBstWouldAddEndPuncttrue
\mciteSetBstMidEndSepPunct{\mcitedefaultmidpunct}
{\mcitedefaultendpunct}{\mcitedefaultseppunct}\relax
\EndOfBibitem
\bibitem[Ahn \latin{et~al.}(2023)Ahn, Triana, Recabal, Herrera, and Simpkins]{ahn2023modification}
Ahn,~W.; Triana,~J.~F.; Recabal,~F.; Herrera,~F.; Simpkins,~B.~S. {Modification of ground-state chemical reactivity via light--matter coherence in infrared cavities}. \emph{Science} \textbf{2023}, \emph{380}, 1165--1168\relax
\mciteBstWouldAddEndPuncttrue
\mciteSetBstMidEndSepPunct{\mcitedefaultmidpunct}
{\mcitedefaultendpunct}{\mcitedefaultseppunct}\relax
\EndOfBibitem
\bibitem[Patrahau \latin{et~al.}(2024)Patrahau, Piejko, Mayer, Antheaume, Sangchai, Ragazzon, Jayachandran, Devaux, Genet, Moran, \latin{et~al.} others]{patrahau2024direct}
Patrahau,~B.; Piejko,~M.; Mayer,~R.~J.; Antheaume,~C.; Sangchai,~T.; Ragazzon,~G.; Jayachandran,~A.; Devaux,~E.; Genet,~C.; Moran,~J.; others {Direct observation of polaritonic chemistry by nuclear magnetic resonance spectroscopy}. \emph{Angew Chem. Int. Ed. Engl.} \textbf{2024}, \emph{63}, e202401368\relax
\mciteBstWouldAddEndPuncttrue
\mciteSetBstMidEndSepPunct{\mcitedefaultmidpunct}
{\mcitedefaultendpunct}{\mcitedefaultseppunct}\relax
\EndOfBibitem
\bibitem[Michon and Simpkins(2024)Michon, and Simpkins]{michon2024impact}
Michon,~M.~A.; Simpkins,~B.~S. {Impact of Cavity Length Non-uniformity on Reaction Rate Extraction in Strong Coupling Experiments}. \emph{J. Am. Chem. Soc.} \textbf{2024}, \emph{146}, 30596--30606\relax
\mciteBstWouldAddEndPuncttrue
\mciteSetBstMidEndSepPunct{\mcitedefaultmidpunct}
{\mcitedefaultendpunct}{\mcitedefaultseppunct}\relax
\EndOfBibitem
\bibitem[Fregoni \latin{et~al.}(2022)Fregoni, Garcia-Vidal, and Feist]{fregoni2022theoretical}
Fregoni,~J.; Garcia-Vidal,~F.~J.; Feist,~J. {Theoretical challenges in polaritonic chemistry}. \emph{ACS photonics} \textbf{2022}, \emph{9}, 1096--1107\relax
\mciteBstWouldAddEndPuncttrue
\mciteSetBstMidEndSepPunct{\mcitedefaultmidpunct}
{\mcitedefaultendpunct}{\mcitedefaultseppunct}\relax
\EndOfBibitem
\bibitem[Foley \latin{et~al.}(2023)Foley, McTague, and DePrince]{foley2023ab}
Foley,~J.~J.; McTague,~J.~F.; DePrince,~A.~E. {Ab initio methods for polariton chemistry}. \emph{Chem. Phys. Rev.} \textbf{2023}, \emph{4}\relax
\mciteBstWouldAddEndPuncttrue
\mciteSetBstMidEndSepPunct{\mcitedefaultmidpunct}
{\mcitedefaultendpunct}{\mcitedefaultseppunct}\relax
\EndOfBibitem
\bibitem[Ruggenthaler \latin{et~al.}(2014)Ruggenthaler, Flick, Pellegrini, Appel, Tokatly, and Rubio]{ruggenthaler2014quantum}
Ruggenthaler,~M.; Flick,~J.; Pellegrini,~C.; Appel,~H.; Tokatly,~I.~V.; Rubio,~A. {Quantum-electrodynamical density-functional theory: Bridging quantum optics and electronic-structure theory}. \emph{Phys. Rev. A} \textbf{2014}, \emph{90}, 012508\relax
\mciteBstWouldAddEndPuncttrue
\mciteSetBstMidEndSepPunct{\mcitedefaultmidpunct}
{\mcitedefaultendpunct}{\mcitedefaultseppunct}\relax
\EndOfBibitem
\bibitem[Ruggenthaler \latin{et~al.}(2018)Ruggenthaler, Tancogne-Dejean, Flick, Appel, and Rubio]{ruggenthaler2018quantum}
Ruggenthaler,~M.; Tancogne-Dejean,~N.; Flick,~J.; Appel,~H.; Rubio,~A. {From a quantum-electrodynamical light--matter description to novel spectroscopies}. \emph{Nat. Rev. Chem.} \textbf{2018}, \emph{2}, 1--16\relax
\mciteBstWouldAddEndPuncttrue
\mciteSetBstMidEndSepPunct{\mcitedefaultmidpunct}
{\mcitedefaultendpunct}{\mcitedefaultseppunct}\relax
\EndOfBibitem
\bibitem[Flick \latin{et~al.}(2018)Flick, Sch\"{a}fer, Ruggenthaler, Appel, and Rubio]{flick2018ab}
Flick,~J.; Sch\"{a}fer,~C.; Ruggenthaler,~M.; Appel,~H.; Rubio,~A. {Ab initio optimized effective potentials for real molecules in optical cavities: Photon contributions to the molecular ground state}. \emph{ACS Photonics} \textbf{2018}, \emph{5}, 992--1005\relax
\mciteBstWouldAddEndPuncttrue
\mciteSetBstMidEndSepPunct{\mcitedefaultmidpunct}
{\mcitedefaultendpunct}{\mcitedefaultseppunct}\relax
\EndOfBibitem
\bibitem[Haugland \latin{et~al.}(2020)Haugland, Ronca, Kj{\o}nstad, Rubio, and Koch]{haugland2020coupled}
Haugland,~T.~S.; Ronca,~E.; Kj{\o}nstad,~E.~F.; Rubio,~A.; Koch,~H. {Coupled cluster theory for molecular polaritons: Changing ground and excited states}. \emph{Phys. Rev. X} \textbf{2020}, \emph{10}, 041043\relax
\mciteBstWouldAddEndPuncttrue
\mciteSetBstMidEndSepPunct{\mcitedefaultmidpunct}
{\mcitedefaultendpunct}{\mcitedefaultseppunct}\relax
\EndOfBibitem
\bibitem[Mordovina \latin{et~al.}(2020)Mordovina, Bungey, Appel, Knowles, Rubio, and Manby]{mordovina2020polaritonic}
Mordovina,~U.; Bungey,~C.; Appel,~H.; Knowles,~P.~J.; Rubio,~A.; Manby,~F.~R. {Polaritonic coupled-cluster theory}. \emph{Phys. Rev. Res.} \textbf{2020}, \emph{2}, 023262\relax
\mciteBstWouldAddEndPuncttrue
\mciteSetBstMidEndSepPunct{\mcitedefaultmidpunct}
{\mcitedefaultendpunct}{\mcitedefaultseppunct}\relax
\EndOfBibitem
\bibitem[Pavosevic and Flick(2021)Pavosevic, and Flick]{pavosevic2021polaritonic}
Pavosevic,~F.; Flick,~J. {Polaritonic unitary coupled cluster for quantum computations}. \emph{J. Phys. Chem. Lett.} \textbf{2021}, \emph{12}, 9100--9107\relax
\mciteBstWouldAddEndPuncttrue
\mciteSetBstMidEndSepPunct{\mcitedefaultmidpunct}
{\mcitedefaultendpunct}{\mcitedefaultseppunct}\relax
\EndOfBibitem
\bibitem[Liebenthal \latin{et~al.}(2022)Liebenthal, Vu, and DePrince]{liebenthal2022equation}
Liebenthal,~M.~D.; Vu,~N.; DePrince,~A.~E. {Equation-of-motion cavity quantum electrodynamics coupled-cluster theory for electron attachment}. \emph{J. Chem. Phys.} \textbf{2022}, \emph{156}\relax
\mciteBstWouldAddEndPuncttrue
\mciteSetBstMidEndSepPunct{\mcitedefaultmidpunct}
{\mcitedefaultendpunct}{\mcitedefaultseppunct}\relax
\EndOfBibitem
\bibitem[Monzel and Stopkowicz(2024)Monzel, and Stopkowicz]{monzel2024diagrams}
Monzel,~L.; Stopkowicz,~S. {Diagrams in Polaritonic Coupled Cluster Theory}. \emph{J. Phys. Chem. A} \textbf{2024}, \emph{128}, 9572--9586\relax
\mciteBstWouldAddEndPuncttrue
\mciteSetBstMidEndSepPunct{\mcitedefaultmidpunct}
{\mcitedefaultendpunct}{\mcitedefaultseppunct}\relax
\EndOfBibitem
\bibitem[Bauer and Dreuw(2023)Bauer, and Dreuw]{bauer2023perturbation}
Bauer,~M.; Dreuw,~A. {Perturbation theoretical approaches to strong light--matter coupling in ground and excited electronic states for the description of molecular polaritons}. \emph{J. Chem. Phys.} \textbf{2023}, \emph{158}\relax
\mciteBstWouldAddEndPuncttrue
\mciteSetBstMidEndSepPunct{\mcitedefaultmidpunct}
{\mcitedefaultendpunct}{\mcitedefaultseppunct}\relax
\EndOfBibitem
\bibitem[El~Moutaoukal \latin{et~al.}(2025)El~Moutaoukal, Riso, Castagnola, Ronca, and Koch]{moutaoukal2025strong}
El~Moutaoukal,~Y.; Riso,~R.~R.; Castagnola,~M.; Ronca,~E.; Koch,~H. {Strong Coupling M{\o}ller--Plesset Perturbation Theory}. \emph{J. Chem. Theory Comput.} \textbf{2025}, \relax
\mciteBstWouldAddEndPunctfalse
\mciteSetBstMidEndSepPunct{\mcitedefaultmidpunct}
{}{\mcitedefaultseppunct}\relax
\EndOfBibitem
\bibitem[Thiam \latin{et~al.}(2024)Thiam, Rossi, Koch, Belpassi, and Ronca]{thiam2024comprehensive}
Thiam,~G.; Rossi,~R.; Koch,~H.; Belpassi,~L.; Ronca,~E. {A comprehensive theory for relativistic polaritonic chemistry: a four components ab initio treatment of molecular systems coupled to quantum fields}. \emph{arXiv preprint arXiv:2409.12757} \textbf{2024}, \relax
\mciteBstWouldAddEndPunctfalse
\mciteSetBstMidEndSepPunct{\mcitedefaultmidpunct}
{}{\mcitedefaultseppunct}\relax
\EndOfBibitem
\bibitem[Konecny \latin{et~al.}(2024)Konecny, Kosheleva, Appel, Ruggenthaler, and Rubio]{konecny2024relativistic}
Konecny,~L.; Kosheleva,~V.~P.; Appel,~H.; Ruggenthaler,~M.; Rubio,~A. Relativistic Linear Response in Quantum-Electrodynamical Density Functional Theory. \emph{arXiv preprint arXiv:2407.02441} \textbf{2024}, \relax
\mciteBstWouldAddEndPunctfalse
\mciteSetBstMidEndSepPunct{\mcitedefaultmidpunct}
{}{\mcitedefaultseppunct}\relax
\EndOfBibitem
\bibitem[Choudhury \latin{et~al.}(2024)Choudhury, Santra, and Ghosh]{choudhury2024understanding}
Choudhury,~A.; Santra,~S.; Ghosh,~D. {Understanding the Photoprocesses in Biological Systems: Need for Accurate Multireference Treatment}. \emph{J. Chem. Theory Comput.} \textbf{2024}, \emph{20}, 4951--4964\relax
\mciteBstWouldAddEndPuncttrue
\mciteSetBstMidEndSepPunct{\mcitedefaultmidpunct}
{\mcitedefaultendpunct}{\mcitedefaultseppunct}\relax
\EndOfBibitem
\bibitem[Lischka \latin{et~al.}(2018)Lischka, Nachtigallova, Aquino, Szalay, Plasser, Machado, and Barbatti]{lischka2018multireference}
Lischka,~H.; Nachtigallova,~D.; Aquino,~A.~J.; Szalay,~P.~G.; Plasser,~F.; Machado,~F.~B.; Barbatti,~M. {Multireference approaches for excited states of molecules}. \emph{Chem. Rev.} \textbf{2018}, \emph{118}, 7293--7361\relax
\mciteBstWouldAddEndPuncttrue
\mciteSetBstMidEndSepPunct{\mcitedefaultmidpunct}
{\mcitedefaultendpunct}{\mcitedefaultseppunct}\relax
\EndOfBibitem
\bibitem[Szalay \latin{et~al.}(2012)Szalay, Muller, Gidofalvi, Lischka, and Shepard]{szalay2012multiconfiguration}
Szalay,~P.~G.; Muller,~T.; Gidofalvi,~G.; Lischka,~H.; Shepard,~R. {Multiconfiguration self-consistent field and multireference configuration interaction methods and applications}. \emph{Chem. Rev.} \textbf{2012}, \emph{112}, 108--181\relax
\mciteBstWouldAddEndPuncttrue
\mciteSetBstMidEndSepPunct{\mcitedefaultmidpunct}
{\mcitedefaultendpunct}{\mcitedefaultseppunct}\relax
\EndOfBibitem
\bibitem[Helgaker \latin{et~al.}(2013)Helgaker, Jorgensen, and Olsen]{helgaker2013molecular}
Helgaker,~T.; Jorgensen,~P.; Olsen,~J. \emph{{Molecular electronic-structure theory}}; John Wiley \& Sons, 2013\relax
\mciteBstWouldAddEndPuncttrue
\mciteSetBstMidEndSepPunct{\mcitedefaultmidpunct}
{\mcitedefaultendpunct}{\mcitedefaultseppunct}\relax
\EndOfBibitem
\bibitem[Vogiatzis \latin{et~al.}(2017)Vogiatzis, Ma, Olsen, Gagliardi, and De~Jong]{vogiatzis2017pushing}
Vogiatzis,~K.~D.; Ma,~D.; Olsen,~J.; Gagliardi,~L.; De~Jong,~W.~A. {Pushing configuration-interaction to the limit: Towards massively parallel MCSCF calculations}. \emph{J. Chem. Phys.} \textbf{2017}, \emph{147}\relax
\mciteBstWouldAddEndPuncttrue
\mciteSetBstMidEndSepPunct{\mcitedefaultmidpunct}
{\mcitedefaultendpunct}{\mcitedefaultseppunct}\relax
\EndOfBibitem
\bibitem[Vu \latin{et~al.}(2024)Vu, Mejia-Rodriguez, Bauman, Panyala, Mutlu, Govind, and Foley~IV]{qed_casci}
Vu,~N.; Mejia-Rodriguez,~D.; Bauman,~N.~P.; Panyala,~A.; Mutlu,~E.; Govind,~N.; Foley~IV,~J.~J. {Cavity Quantum Electrodynamics Complete Active Space Configuration Interaction Theory}. \emph{J. Chem. Theory Comput.} \textbf{2024}, \emph{20}, 1214--1227\relax
\mciteBstWouldAddEndPuncttrue
\mciteSetBstMidEndSepPunct{\mcitedefaultmidpunct}
{\mcitedefaultendpunct}{\mcitedefaultseppunct}\relax
\EndOfBibitem
\bibitem[Matou\v{s}ek \latin{et~al.}(2024)Matou\v{s}ek, Vu, Govind, Foley~IV, and Veis]{qed_dmrg}
Matou\v{s}ek,~M.; Vu,~N.; Govind,~N.; Foley~IV,~J.~J.; Veis,~L. {Polaritonic chemistry using the density matrix renormalization group method}. \emph{J. Chem. Theory Comput.} \textbf{2024}, \emph{20}, 9424--9434\relax
\mciteBstWouldAddEndPuncttrue
\mciteSetBstMidEndSepPunct{\mcitedefaultmidpunct}
{\mcitedefaultendpunct}{\mcitedefaultseppunct}\relax
\EndOfBibitem
\bibitem[Mallory and DePrince~III(2022)Mallory, and DePrince~III]{qed_rmd2}
Mallory,~J.~D.; DePrince~III,~A.~E. {Reduced-density-matrix-based ab initio cavity quantum electrodynamics}. \emph{Phys. Rev. A} \textbf{2022}, \emph{106}, 053710\relax
\mciteBstWouldAddEndPuncttrue
\mciteSetBstMidEndSepPunct{\mcitedefaultmidpunct}
{\mcitedefaultendpunct}{\mcitedefaultseppunct}\relax
\EndOfBibitem
\bibitem[Levine \latin{et~al.}(2021)Levine, Durden, Esch, Liang, and Shu]{levine2021cas}
Levine,~B.~G.; Durden,~A.~S.; Esch,~M.~P.; Liang,~F.; Shu,~Y. {CAS without SCF—Why to use CASCI and where to get the orbitals}. \emph{J. Chem. Phys.} \textbf{2021}, \emph{154}\relax
\mciteBstWouldAddEndPuncttrue
\mciteSetBstMidEndSepPunct{\mcitedefaultmidpunct}
{\mcitedefaultendpunct}{\mcitedefaultseppunct}\relax
\EndOfBibitem
\bibitem[Jensen and J{\o}rgensen(1984)Jensen, and J{\o}rgensen]{jensen1984direct}
Jensen,~H. J.~A.; J{\o}rgensen,~P. {A direct approach to second-order MCSCF calculations using a norm extended optimization scheme}. \emph{J. Chem. Phys.} \textbf{1984}, \emph{80}, 1204--1214\relax
\mciteBstWouldAddEndPuncttrue
\mciteSetBstMidEndSepPunct{\mcitedefaultmidpunct}
{\mcitedefaultendpunct}{\mcitedefaultseppunct}\relax
\EndOfBibitem
\bibitem[H{\o}yvik \latin{et~al.}(2012)H{\o}yvik, Jansik, and J{\o}rgensen]{hoyvik2012trust}
H{\o}yvik,~I.-M.; Jansik,~B.; J{\o}rgensen,~P. {Trust region minimization of orbital localization functions}. \emph{J. Chem. Theory Comput.} \textbf{2012}, \emph{8}, 3137--3146\relax
\mciteBstWouldAddEndPuncttrue
\mciteSetBstMidEndSepPunct{\mcitedefaultmidpunct}
{\mcitedefaultendpunct}{\mcitedefaultseppunct}\relax
\EndOfBibitem
\bibitem[Folkestad \latin{et~al.}(2022)Folkestad, Matveeva, H{\o}yvik, and Koch]{folkestad2022implementation}
Folkestad,~S.~D.; Matveeva,~R.; H{\o}yvik,~I.-M.; Koch,~H. {Implementation of occupied and virtual Edmiston--Ruedenberg orbitals using Cholesky decomposed integrals}. \emph{J. Chem. Theory Comput.} \textbf{2022}, \emph{18}, 4733--4744\relax
\mciteBstWouldAddEndPuncttrue
\mciteSetBstMidEndSepPunct{\mcitedefaultmidpunct}
{\mcitedefaultendpunct}{\mcitedefaultseppunct}\relax
\EndOfBibitem
\bibitem[Helmich-Paris(2022)]{helmich2022trust}
Helmich-Paris,~B. {A trust-region augmented Hessian implementation for state-specific and state-averaged CASSCF wave functions}. \emph{J. Chem. Phys.} \textbf{2022}, \emph{156}\relax
\mciteBstWouldAddEndPuncttrue
\mciteSetBstMidEndSepPunct{\mcitedefaultmidpunct}
{\mcitedefaultendpunct}{\mcitedefaultseppunct}\relax
\EndOfBibitem
\bibitem[Vu \latin{et~al.}(2025)Vu, Ampoh, and Foley]{vu2025cavity}
Vu,~N.; Ampoh,~K.; Foley,~J. {Cavity quantum electrodynamics complete active space self consistent field theory}. \emph{ChemRxiv preprint doi:10.26434/chemrxiv-2025-q6rfm} \textbf{2025}, \relax
\mciteBstWouldAddEndPunctfalse
\mciteSetBstMidEndSepPunct{\mcitedefaultmidpunct}
{}{\mcitedefaultseppunct}\relax
\EndOfBibitem
\bibitem[Olsen \latin{et~al.}(1988)Olsen, Roos, Jo/rgensen, and Jensen]{olsen1988determinant}
Olsen,~J.; Roos,~B.~O.; Jo/rgensen,~P.; Jensen,~H. J.~A. {Determinant based configuration interaction algorithms for complete and restricted configuration interaction spaces}. \emph{J. Chem. Phys.} \textbf{1988}, \emph{89}, 2185--2192\relax
\mciteBstWouldAddEndPuncttrue
\mciteSetBstMidEndSepPunct{\mcitedefaultmidpunct}
{\mcitedefaultendpunct}{\mcitedefaultseppunct}\relax
\EndOfBibitem
\bibitem[Malmqvist \latin{et~al.}(1990)Malmqvist, Rendell, and Roos]{malmqvist1990restricted}
Malmqvist,~P.~{\AA}.; Rendell,~A.; Roos,~B.~O. {The restricted active space self-consistent-field method, implemented with a split graph unitary group approach}. \emph{J. Phys. Chem.} \textbf{1990}, \emph{94}, 5477--5482\relax
\mciteBstWouldAddEndPuncttrue
\mciteSetBstMidEndSepPunct{\mcitedefaultmidpunct}
{\mcitedefaultendpunct}{\mcitedefaultseppunct}\relax
\EndOfBibitem
\bibitem[Fleig \latin{et~al.}(2001)Fleig, Olsen, and Marian]{fleig2001generalized}
Fleig,~T.; Olsen,~J.; Marian,~C.~M. {The generalized active space concept for the relativistic treatment of electron correlation. I. Kramers-restricted two-component configuration interaction}. \emph{J. Chem. Phys.} \textbf{2001}, \emph{114}, 4775--4790\relax
\mciteBstWouldAddEndPuncttrue
\mciteSetBstMidEndSepPunct{\mcitedefaultmidpunct}
{\mcitedefaultendpunct}{\mcitedefaultseppunct}\relax
\EndOfBibitem
\bibitem[Schollw{\"o}ck(2005)]{schollwock2005density}
Schollw{\"o}ck,~U. {The density-matrix renormalization group}. \emph{Rev. Mod. Phys.} \textbf{2005}, \emph{77}, 259--315\relax
\mciteBstWouldAddEndPuncttrue
\mciteSetBstMidEndSepPunct{\mcitedefaultmidpunct}
{\mcitedefaultendpunct}{\mcitedefaultseppunct}\relax
\EndOfBibitem
\bibitem[Mazziotti(2006)]{mazziotti2006quantum}
Mazziotti,~D.~A. {Quantum chemistry without wave functions: Two-electron reduced density matrices}. \emph{Acc. Chem. Res.} \textbf{2006}, \emph{39}, 207--215\relax
\mciteBstWouldAddEndPuncttrue
\mciteSetBstMidEndSepPunct{\mcitedefaultmidpunct}
{\mcitedefaultendpunct}{\mcitedefaultseppunct}\relax
\EndOfBibitem
\bibitem[Roos \latin{et~al.}(1980)Roos, Taylor, and Sigbahn]{roos1980sci1}
Roos,~B.~O.; Taylor,~P.~R.; Sigbahn,~P.~E. {A complete active space SCF method (CASSCF) using a density matrix formulated super-CI approach}. \emph{Chem. Phys.} \textbf{1980}, \emph{48}, 157--173\relax
\mciteBstWouldAddEndPuncttrue
\mciteSetBstMidEndSepPunct{\mcitedefaultmidpunct}
{\mcitedefaultendpunct}{\mcitedefaultseppunct}\relax
\EndOfBibitem
\bibitem[Roos(1980)]{roos1980sci2}
Roos,~B.~O. {The complete active space SCF method in a fock-matrix-based super-CI formulation}. \emph{Int. J. Quantum. Chem.} \textbf{1980}, \emph{18}, 175--189\relax
\mciteBstWouldAddEndPuncttrue
\mciteSetBstMidEndSepPunct{\mcitedefaultmidpunct}
{\mcitedefaultendpunct}{\mcitedefaultseppunct}\relax
\EndOfBibitem
\bibitem[Kollmar \latin{et~al.}(2019)Kollmar, Sivalingam, Helmich-Paris, Angeli, and Neese]{superCI_PT}
Kollmar,~C.; Sivalingam,~K.; Helmich-Paris,~B.; Angeli,~C.; Neese,~F. {A perturbation-based super-CI approach for the orbital optimization of a CASSCF wave function}. \emph{J. Comput. Chem.} \textbf{2019}, \emph{40}, 1463--1470\relax
\mciteBstWouldAddEndPuncttrue
\mciteSetBstMidEndSepPunct{\mcitedefaultmidpunct}
{\mcitedefaultendpunct}{\mcitedefaultseppunct}\relax
\EndOfBibitem
\bibitem[Jensen and {\AA}gren(1986)Jensen, and {\AA}gren]{jensen1986direct}
Jensen,~H. J.~A.; {\AA}gren,~H. {A direct, restricted-step, second-order MC SCF program for large scale ab initio calculations}. \emph{Chem. Phys.} \textbf{1986}, \emph{104}, 229--250\relax
\mciteBstWouldAddEndPuncttrue
\mciteSetBstMidEndSepPunct{\mcitedefaultmidpunct}
{\mcitedefaultendpunct}{\mcitedefaultseppunct}\relax
\EndOfBibitem
\bibitem[Jensen \latin{et~al.}(1987)Jensen, J{\o}rgensen, and {\AA}gren]{jensen1987efficient}
Jensen,~H. J.~A.; J{\o}rgensen,~P.; {\AA}gren,~H. {Efficient optimization of large scale MCSCF wave functions with a restricted step algorithm}. \emph{J. Chem. Phys.} \textbf{1987}, \emph{87}, 451--466\relax
\mciteBstWouldAddEndPuncttrue
\mciteSetBstMidEndSepPunct{\mcitedefaultmidpunct}
{\mcitedefaultendpunct}{\mcitedefaultseppunct}\relax
\EndOfBibitem
\bibitem[Werner and Meyer(1980)Werner, and Meyer]{werner1980quadratically}
Werner,~H.-J.; Meyer,~W. {A quadratically convergent multiconfiguration--self-consistent field method with simultaneous optimization of orbitals and CI coefficients}. \emph{J. Chem. Phys.} \textbf{1980}, \emph{73}, 2342--2356\relax
\mciteBstWouldAddEndPuncttrue
\mciteSetBstMidEndSepPunct{\mcitedefaultmidpunct}
{\mcitedefaultendpunct}{\mcitedefaultseppunct}\relax
\EndOfBibitem
\bibitem[Werner and Knowles(1985)Werner, and Knowles]{werner1985second}
Werner,~H.-J.; Knowles,~P.~J. {A second order multiconfiguration SCF procedure with optimum convergence}. \emph{J. Chem. Phys.} \textbf{1985}, \emph{82}, 5053--5063\relax
\mciteBstWouldAddEndPuncttrue
\mciteSetBstMidEndSepPunct{\mcitedefaultmidpunct}
{\mcitedefaultendpunct}{\mcitedefaultseppunct}\relax
\EndOfBibitem
\bibitem[Sun \latin{et~al.}(2017)Sun, Yang, and Chan]{sun2017general}
Sun,~Q.; Yang,~J.; Chan,~G. K.-L. {A general second order complete active space self-consistent-field solver for large-scale systems}. \emph{Chem. Phys. Lett.} \textbf{2017}, \emph{683}, 291--299\relax
\mciteBstWouldAddEndPuncttrue
\mciteSetBstMidEndSepPunct{\mcitedefaultmidpunct}
{\mcitedefaultendpunct}{\mcitedefaultseppunct}\relax
\EndOfBibitem
\bibitem[Ma \latin{et~al.}(2017)Ma, Knecht, Keller, and Reiher]{ma2017second}
Ma,~Y.; Knecht,~S.; Keller,~S.; Reiher,~M. {Second-order self-consistent-field density-matrix renormalization group}. \emph{J. Chem. Theory Comput.} \textbf{2017}, \emph{13}, 2533--2549\relax
\mciteBstWouldAddEndPuncttrue
\mciteSetBstMidEndSepPunct{\mcitedefaultmidpunct}
{\mcitedefaultendpunct}{\mcitedefaultseppunct}\relax
\EndOfBibitem
\bibitem[Kreplin \latin{et~al.}(2019)Kreplin, Knowles, and Werner]{kreplin2019second}
Kreplin,~D.~A.; Knowles,~P.~J.; Werner,~H.-J. {Second-order MCSCF optimization revisited. I. Improved algorithms for fast and robust second-order CASSCF convergence}. \emph{J. Chem. Phys.} \textbf{2019}, \emph{150}\relax
\mciteBstWouldAddEndPuncttrue
\mciteSetBstMidEndSepPunct{\mcitedefaultmidpunct}
{\mcitedefaultendpunct}{\mcitedefaultseppunct}\relax
\EndOfBibitem
\bibitem[Kreplin \latin{et~al.}(2020)Kreplin, Knowles, and Werner]{kreplin2020mcscf}
Kreplin,~D.~A.; Knowles,~P.~J.; Werner,~H.-J. {MCSCF optimization revisited. II. Combined first-and second-order orbital optimization for large molecules}. \emph{J. Chem. Phys.} \textbf{2020}, \emph{152}\relax
\mciteBstWouldAddEndPuncttrue
\mciteSetBstMidEndSepPunct{\mcitedefaultmidpunct}
{\mcitedefaultendpunct}{\mcitedefaultseppunct}\relax
\EndOfBibitem
\bibitem[Nottoli \latin{et~al.}(2021)Nottoli, Gauss, and Lipparini]{nottoli2021second}
Nottoli,~T.; Gauss,~J.; Lipparini,~F. {Second-order CASSCF algorithm with the Cholesky decomposition of the two-electron integrals}. \emph{J. Chem. Theory Comput.} \textbf{2021}, \emph{17}, 6819--6831\relax
\mciteBstWouldAddEndPuncttrue
\mciteSetBstMidEndSepPunct{\mcitedefaultmidpunct}
{\mcitedefaultendpunct}{\mcitedefaultseppunct}\relax
\EndOfBibitem
\bibitem[Aidas \latin{et~al.}(2014)Aidas, Angeli, Bak, Bakken, Bast, Boman, Christiansen, Cimiraglia, Coriani, Dahle, \latin{et~al.} others]{aidas2014d}
Aidas,~K.; Angeli,~C.; Bak,~K.~L.; Bakken,~V.; Bast,~R.; Boman,~L.; Christiansen,~O.; Cimiraglia,~R.; Coriani,~S.; Dahle,~P.; others {The Dalton quantum chemistry program system}. \emph{Wiley Interdiscip. Rev. Comput. Mol. Sci.} \textbf{2014}, \emph{4}, 269--284\relax
\mciteBstWouldAddEndPuncttrue
\mciteSetBstMidEndSepPunct{\mcitedefaultmidpunct}
{\mcitedefaultendpunct}{\mcitedefaultseppunct}\relax
\EndOfBibitem
\bibitem[Aquilante \latin{et~al.}(2020)Aquilante, Autschbach, Baiardi, Battaglia, Borin, Chibotaru, Conti, De~Vico, Delcey, Ferr{\'e}, \latin{et~al.} others]{aquilante2020modern}
Aquilante,~F.; Autschbach,~J.; Baiardi,~A.; Battaglia,~S.; Borin,~V.~A.; Chibotaru,~L.~F.; Conti,~I.; De~Vico,~L.; Delcey,~M.; Ferr{\'e},~N.; others {Modern quantum chemistry with [Open] Molcas}. \emph{J. Chem. Phys.} \textbf{2020}, \emph{152}\relax
\mciteBstWouldAddEndPuncttrue
\mciteSetBstMidEndSepPunct{\mcitedefaultmidpunct}
{\mcitedefaultendpunct}{\mcitedefaultseppunct}\relax
\EndOfBibitem
\bibitem[Matthews \latin{et~al.}(2020)Matthews, Cheng, Harding, Lipparini, Stopkowicz, Jagau, Szalay, Gauss, and Stanton]{matthews2020coupled}
Matthews,~D.~A.; Cheng,~L.; Harding,~M.~E.; Lipparini,~F.; Stopkowicz,~S.; Jagau,~T.-C.; Szalay,~P.~G.; Gauss,~J.; Stanton,~J.~F. {Coupled-cluster techniques for computational chemistry: The CFOUR program package}. \emph{J. Chem. Phys.} \textbf{2020}, \emph{152}\relax
\mciteBstWouldAddEndPuncttrue
\mciteSetBstMidEndSepPunct{\mcitedefaultmidpunct}
{\mcitedefaultendpunct}{\mcitedefaultseppunct}\relax
\EndOfBibitem
\bibitem[Sun \latin{et~al.}(2020)Sun, Zhang, Banerjee, Bao, Barbry, Blunt, Bogdanov, Booth, Chen, Cui, \latin{et~al.} others]{sun2020recent}
Sun,~Q.; Zhang,~X.; Banerjee,~S.; Bao,~P.; Barbry,~M.; Blunt,~N.~S.; Bogdanov,~N.~A.; Booth,~G.~H.; Chen,~J.; Cui,~Z.-H.; others {Recent developments in the PySCF program package}. \emph{J. Chem. Phys.} \textbf{2020}, \emph{153}\relax
\mciteBstWouldAddEndPuncttrue
\mciteSetBstMidEndSepPunct{\mcitedefaultmidpunct}
{\mcitedefaultendpunct}{\mcitedefaultseppunct}\relax
\EndOfBibitem
\bibitem[Neese(2012)]{neese2012orca}
Neese,~F. {The ORCA program system}. \emph{Wiley Interdiscip. Rev. Comput. Mol. Sci.} \textbf{2012}, \emph{2}, 73--78\relax
\mciteBstWouldAddEndPuncttrue
\mciteSetBstMidEndSepPunct{\mcitedefaultmidpunct}
{\mcitedefaultendpunct}{\mcitedefaultseppunct}\relax
\EndOfBibitem
\bibitem[Davidson(1975)]{davidson197514}
Davidson,~E.~R. {The Iterative Calculation of a Few of the Lowest Eigenvalues and Corresponding Eigenvectors of Large Real-Symmetric Matrices}. \emph{J. Comput. Phys.} \textbf{1975}, \emph{17}, 87--94\relax
\mciteBstWouldAddEndPuncttrue
\mciteSetBstMidEndSepPunct{\mcitedefaultmidpunct}
{\mcitedefaultendpunct}{\mcitedefaultseppunct}\relax
\EndOfBibitem
\bibitem[Roos(1972)]{roos1972new}
Roos,~B. {A new method for large-scale Cl calculations}. \emph{Chem. Phys. Lett.} \textbf{1972}, \emph{15}, 153--159\relax
\mciteBstWouldAddEndPuncttrue
\mciteSetBstMidEndSepPunct{\mcitedefaultmidpunct}
{\mcitedefaultendpunct}{\mcitedefaultseppunct}\relax
\EndOfBibitem
\bibitem[Fletcher(2000)]{fletcher2000practical}
Fletcher,~R. \emph{{Practical methods of optimization}}; John Wiley \& Sons, 2000\relax
\mciteBstWouldAddEndPuncttrue
\mciteSetBstMidEndSepPunct{\mcitedefaultmidpunct}
{\mcitedefaultendpunct}{\mcitedefaultseppunct}\relax
\EndOfBibitem
\bibitem[Simons \latin{et~al.}(1983)Simons, Joergensen, Taylor, and Ozment]{simons1983walking}
Simons,~J.; Joergensen,~P.; Taylor,~H.; Ozment,~J. {Walking on potential energy surfaces}. \emph{J. Phys. Chem.} \textbf{1983}, \emph{87}, 2745--2753\relax
\mciteBstWouldAddEndPuncttrue
\mciteSetBstMidEndSepPunct{\mcitedefaultmidpunct}
{\mcitedefaultendpunct}{\mcitedefaultseppunct}\relax
\EndOfBibitem
\bibitem[Levenberg(1944)]{levenberg1944method}
Levenberg,~K. {A method for the solution of certain non-linear problems in least squares}. \emph{Q. Appl. Math.} \textbf{1944}, \emph{2}, 164--168\relax
\mciteBstWouldAddEndPuncttrue
\mciteSetBstMidEndSepPunct{\mcitedefaultmidpunct}
{\mcitedefaultendpunct}{\mcitedefaultseppunct}\relax
\EndOfBibitem
\bibitem[Marquardt(1963)]{marquardt1963algorithm}
Marquardt,~D.~W. An algorithm for least-squares estimation of nonlinear parameters. \emph{SIAM} \textbf{1963}, \emph{11}, 431--441\relax
\mciteBstWouldAddEndPuncttrue
\mciteSetBstMidEndSepPunct{\mcitedefaultmidpunct}
{\mcitedefaultendpunct}{\mcitedefaultseppunct}\relax
\EndOfBibitem
\bibitem[Hylleraas and Undheim(1930)Hylleraas, and Undheim]{hylleraas1930numerische}
Hylleraas,~E.~A.; Undheim,~B. {Numerische berechnung der 2 S-terme von ortho-und par-helium}. \emph{Z. Phys.} \textbf{1930}, \emph{65}, 759--772\relax
\mciteBstWouldAddEndPuncttrue
\mciteSetBstMidEndSepPunct{\mcitedefaultmidpunct}
{\mcitedefaultendpunct}{\mcitedefaultseppunct}\relax
\EndOfBibitem
\bibitem[MacDonald(1933)]{macdonald1933successive}
MacDonald,~J. {Successive approximations by the Rayleigh-Ritz variation method}. \emph{Phys. Rev.} \textbf{1933}, \emph{43}, 830\relax
\mciteBstWouldAddEndPuncttrue
\mciteSetBstMidEndSepPunct{\mcitedefaultmidpunct}
{\mcitedefaultendpunct}{\mcitedefaultseppunct}\relax
\EndOfBibitem
\bibitem[Folkestad \latin{et~al.}(2020)Folkestad, Kj{\o}nstad, Myhre, Andersen, Balbi, Coriani, Giovannini, Goletto, Haugland, Hutcheson, \latin{et~al.} others]{eT}
Folkestad,~S.~D.; Kj{\o}nstad,~E.~F.; Myhre,~R.~H.; Andersen,~J.~H.; Balbi,~A.; Coriani,~S.; Giovannini,~T.; Goletto,~L.; Haugland,~T.~S.; Hutcheson,~A.; others {eT 1.0: An open source electronic structure program with emphasis on coupled cluster and multilevel methods}. \emph{J. Chem. Phys.} \textbf{2020}, \emph{152}\relax
\mciteBstWouldAddEndPuncttrue
\mciteSetBstMidEndSepPunct{\mcitedefaultmidpunct}
{\mcitedefaultendpunct}{\mcitedefaultseppunct}\relax
\EndOfBibitem
\bibitem[Lexander \latin{et~al.}(2024)Lexander, Angelico, Kj{\o}nstad, and Koch]{lexander2024analytical}
Lexander,~M.~T.; Angelico,~S.; Kj{\o}nstad,~E.~F.; Koch,~H. {Analytical evaluation of ground state gradients in quantum electrodynamics coupled cluster theory}. \emph{J. Chem. Theory Comput.} \textbf{2024}, \emph{20}, 8876--8885\relax
\mciteBstWouldAddEndPuncttrue
\mciteSetBstMidEndSepPunct{\mcitedefaultmidpunct}
{\mcitedefaultendpunct}{\mcitedefaultseppunct}\relax
\EndOfBibitem
\bibitem[Dunning~Jr(1989)]{dunning1989gaussian}
Dunning~Jr,~T.~H. {Gaussian basis sets for use in correlated molecular calculations. I. The atoms boron through neon and hydrogen}. \emph{J. Chem. Phys.} \textbf{1989}, \emph{90}, 1007--1023\relax
\mciteBstWouldAddEndPuncttrue
\mciteSetBstMidEndSepPunct{\mcitedefaultmidpunct}
{\mcitedefaultendpunct}{\mcitedefaultseppunct}\relax
\EndOfBibitem
\bibitem[Mai and Gonz{\'a}lez(2020)Mai, and Gonz{\'a}lez]{mai2020molecular}
Mai,~S.; Gonz{\'a}lez,~L. {Molecular photochemistry: recent developments in theory}. \emph{Angew. Chem. Int. Ed. Eng.} \textbf{2020}, \emph{59}, 16832--16846\relax
\mciteBstWouldAddEndPuncttrue
\mciteSetBstMidEndSepPunct{\mcitedefaultmidpunct}
{\mcitedefaultendpunct}{\mcitedefaultseppunct}\relax
\EndOfBibitem
\bibitem[Feller \latin{et~al.}(2014)Feller, Peterson, and Davidson]{feller2014systematic}
Feller,~D.; Peterson,~K.~A.; Davidson,~E.~R. {A systematic approach to vertically excited states of ethylene using configuration interaction and coupled cluster techniques}. \emph{J. Chem. Phys.} \textbf{2014}, \emph{141}, 104302\relax
\mciteBstWouldAddEndPuncttrue
\mciteSetBstMidEndSepPunct{\mcitedefaultmidpunct}
{\mcitedefaultendpunct}{\mcitedefaultseppunct}\relax
\EndOfBibitem
\bibitem[Riso \latin{et~al.}(2022)Riso, Haugland, Ronca, and Koch]{riso2022molecular}
Riso,~R.~R.; Haugland,~T.~S.; Ronca,~E.; Koch,~H. {Molecular orbital theory in cavity QED environments}. \emph{Nat. Comm.} \textbf{2022}, \emph{13}, 1368\relax
\mciteBstWouldAddEndPuncttrue
\mciteSetBstMidEndSepPunct{\mcitedefaultmidpunct}
{\mcitedefaultendpunct}{\mcitedefaultseppunct}\relax
\EndOfBibitem
\bibitem[El~Moutaoukal \latin{et~al.}(2024)El~Moutaoukal, Riso, Castagnola, and Koch]{el2024toward}
El~Moutaoukal,~Y.; Riso,~R.~R.; Castagnola,~M.; Koch,~H. {Toward polaritonic molecular orbitals for large molecular systems}. \emph{J. Chem. Theory Comput.} \textbf{2024}, \emph{20}, 8911--8920\relax
\mciteBstWouldAddEndPuncttrue
\mciteSetBstMidEndSepPunct{\mcitedefaultmidpunct}
{\mcitedefaultendpunct}{\mcitedefaultseppunct}\relax
\EndOfBibitem
\bibitem[Castagnola \latin{et~al.}(2025)Castagnola, Riso, El~Moutaoukal, Ronca, and Koch]{castagnola2025strong}
Castagnola,~M.; Riso,~R.~R.; El~Moutaoukal,~Y.; Ronca,~E.; Koch,~H. {Strong Coupling Quantum Electrodynamics Hartree--Fock Response Theory}. \emph{J. Phys. Chem. A} \textbf{2025}, \relax
\mciteBstWouldAddEndPunctfalse
\mciteSetBstMidEndSepPunct{\mcitedefaultmidpunct}
{}{\mcitedefaultseppunct}\relax
\EndOfBibitem
\bibitem[Castagnola \latin{et~al.}(2025)Castagnola, Lexander, and Koch]{castagnola2024realistic}
Castagnola,~M.; Lexander,~M.~T.; Koch,~H. {Realistic ab initio predictions of excimer behavior under collective light-matter strong coupling}. \emph{Phys. Rev. X} \textbf{2025}, \emph{15}, 021040\relax
\mciteBstWouldAddEndPuncttrue
\mciteSetBstMidEndSepPunct{\mcitedefaultmidpunct}
{\mcitedefaultendpunct}{\mcitedefaultseppunct}\relax
\EndOfBibitem
\bibitem[Koessler \latin{et~al.}(2025)Koessler, Mandal, Musser, Krauss, and Huo]{koessler2025polariton}
Koessler,~E.; Mandal,~A.; Musser,~A.; Krauss,~T.; Huo,~P. {Polariton Mediated Electron Transfer under the Collective Molecule-Cavity Coupling Regime}. \emph{ChemRxiv preprint doi:10.26434/chemrxiv-2025-s1r54} \textbf{2025}, \relax
\mciteBstWouldAddEndPunctfalse
\mciteSetBstMidEndSepPunct{\mcitedefaultmidpunct}
{}{\mcitedefaultseppunct}\relax
\EndOfBibitem
\bibitem[Horak \latin{et~al.}(2025)Horak, Sidler, Schnappinger, Huang, Ruggenthaler, and Rubio]{horak2025analytic}
Horak,~J.; Sidler,~D.; Schnappinger,~T.; Huang,~W.-M.; Ruggenthaler,~M.; Rubio,~A. {Analytic model reveals local molecular polarizability changes induced by collective strong coupling in optical cavities}. \emph{Phys. Rev. Res.} \textbf{2025}, \emph{7}, 013242\relax
\mciteBstWouldAddEndPuncttrue
\mciteSetBstMidEndSepPunct{\mcitedefaultmidpunct}
{\mcitedefaultendpunct}{\mcitedefaultseppunct}\relax
\EndOfBibitem
\bibitem[Galego \latin{et~al.}(2015)Galego, Garcia-Vidal, and Feist]{galego2015cavity}
Galego,~J.; Garcia-Vidal,~F.~J.; Feist,~J. {Cavity-induced modifications of molecular structure in the strong-coupling regime}. \emph{Phys. Rev. X} \textbf{2015}, \emph{5}, 041022\relax
\mciteBstWouldAddEndPuncttrue
\mciteSetBstMidEndSepPunct{\mcitedefaultmidpunct}
{\mcitedefaultendpunct}{\mcitedefaultseppunct}\relax
\EndOfBibitem
\bibitem[Bennett \latin{et~al.}(2016)Bennett, Kowalewski, and Mukamel]{bennett2016novel}
Bennett,~K.; Kowalewski,~M.; Mukamel,~S. {Novel photochemistry of molecular polaritons in optical cavities}. \emph{Faraday Discuss.} \textbf{2016}, \emph{194}, 259--282\relax
\mciteBstWouldAddEndPuncttrue
\mciteSetBstMidEndSepPunct{\mcitedefaultmidpunct}
{\mcitedefaultendpunct}{\mcitedefaultseppunct}\relax
\EndOfBibitem
\bibitem[Rana \latin{et~al.}(2023)Rana, Hohenstein, and Mart{\'\i}nez]{rana2023simulating}
Rana,~B.; Hohenstein,~E.~G.; Mart{\'\i}nez,~T.~J. {Simulating the excited-state dynamics of polaritons with ab initio multiple spawning}. \emph{J. Phys. Chem. A} \textbf{2023}, \emph{128}, 139--151\relax
\mciteBstWouldAddEndPuncttrue
\mciteSetBstMidEndSepPunct{\mcitedefaultmidpunct}
{\mcitedefaultendpunct}{\mcitedefaultseppunct}\relax
\EndOfBibitem
\end{mcitethebibliography}

\providecommand{\latin}[1]{#1}
\makeatletter
\providecommand{\doi}
  {\begingroup\let\do\@makeother\dospecials
  \catcode`\{=1 \catcode`\}=2 \doi@aux}
\providecommand{\doi@aux}[1]{\endgroup\texttt{#1}}
\makeatother
\providecommand*\mcitethebibliography{\thebibliography}
\csname @ifundefined\endcsname{endmcitethebibliography}  {\let\endmcitethebibliography\endthebibliography}{}

\includepdf[pages=-]{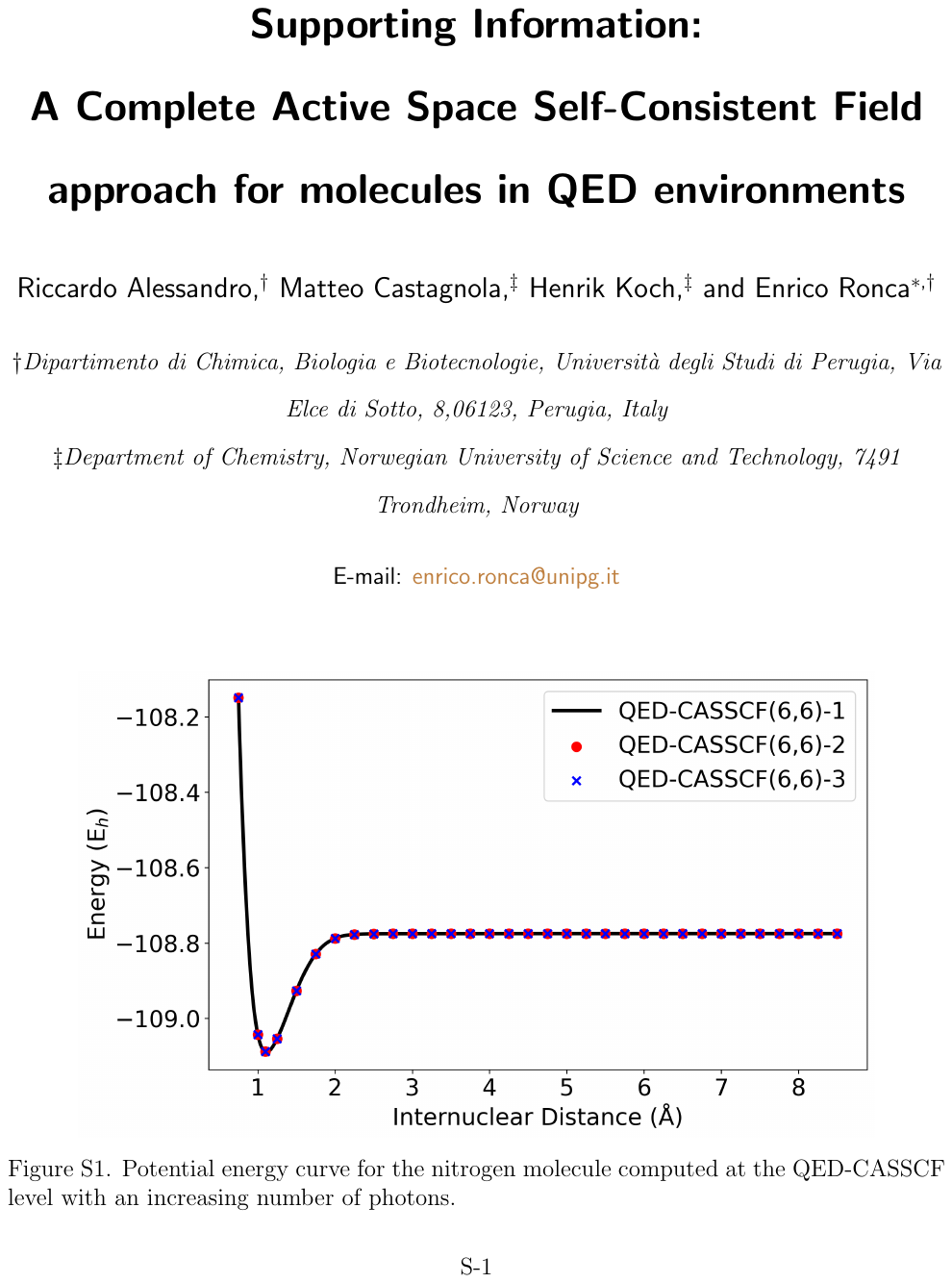}

\end{document}